\documentclass{statsoc}

\usepackage[a4paper]{geometry}
\usepackage[textsize=small]{todonotes}
\usepackage{graphicx}
\usepackage{amsmath}
\DeclareMathOperator*{\argmax}{arg\,max}
\usepackage{amssymb}
\usepackage{enumerate}
\usepackage{comment}
\usepackage{float}
\usepackage{natbib}
\usepackage[caption=false]{subfig}
\newtheorem{remark}{Remark}
\usepackage{hyperref}

\usepackage{etoolbox}
\makeatletter
\patchcmd{\@makecaption}
  {\parbox}
  {\advance\@tempdima-\fontdimen2}
  {}{}
\makeatother  

\title[Outcome-guided Sparse K-means]{Outcome-guided Sparse K-means for Disease Subtype Discovery via Integrating Phenotypic Data with High-dimensional Transcriptomic Data}
\author[Author 1 {\it et al.}]{Lingsong Meng}
\address{Department of Biostatistics, University of Florida, Gainesville, USA.}
\author{Dorina Avram}
\address{Department of Immunology, H. Lee Moffitt Cancer Center and Research Institute, Tampa, USA.}
\author{George Tseng}
\address{Department of Biostatistics, University of Pittsburgh, Pittsburgh, USA.}
\author[L.Meng, D.Avram, G.Tseng and Z.Huo]{and Zhiguang Huo} \coaddress{Zhiguang Huo, Department of Biostatistics, University of Florida, 2004 Mowry Road, Gainesville, FL 32611-7450, USA\\ E-mail: zhuo@ufl.edu}
\address{Department of Biostatistics, University of Florida, Gainesville, USA.}

\begin{document}
\begin{abstract}
The discovery of disease subtypes is an essential step for developing precision medicine, and disease subtyping via omics data has become a popular approach. 
While promising, subtypes obtained from existing approaches are not necessarily associated with clinical outcomes. 
With the rich clinical data along with the omics data in modern epidemiology cohorts, it is urgent to develop an outcome-guided clustering algorithm to fully integrate the phenotypic data with the high-dimensional omics data. 
Hence, we extended a sparse K-means method to an outcome-guided sparse K-means (GuidedSparseKmeans) method. 
An unified objective function was proposed, which was comprised of (i) weighted K-means to perform sample clusterings; (ii) lasso regularizations to perform gene selection from the high-dimensional omics data; (iii) incorporation of a phenotypic variable from the clinical dataset to facilitate biologically meaningful clustering results.
By iteratively optimizing the objective function, we will simultaneously obtain a phenotype-related sample clustering results and gene selection results.
We demonstrated the superior performance of the GuidedSparseKmeans by comparing with existing clustering methods in simulations and applications of high-dimensional transcriptomic data of breast cancer and Alzheimer’s disease. Our algorithm has been implemented into an R package, which is publicly available on GitHub (\url{https://github.com/LingsongMeng/GuidedSparseKmeans}).

\textit{Keywords}: Outcome-guided clustering; sparse K-means; Disease subtype; Phenotypic data; Transcriptomic data 
\end{abstract}

\section{Introduction}
\label{sec:intro}

Historically, 
many complex diseases were thought of as a single disease, 
but modern omics studies have revealed that a single disease can be heterogeneous and contain several disease subtypes.
Identification of these disease subtypes is an essential step toward precision medicine, 
since different subtypes usually show distinct clinical responses to different treatments \citep{abramson2015subtyping}.
Disease subtyping via omics data has been prevalent in the literature, 
and representative diseases include 
lymphoma \citep{rosenwald2002use},
glioblastoma \citep{parsons2008integrated, verhaak2010integrated},
breast cancer \citep{lehmann2011identification, parker2009supervised},
colorectal cancer \citep{sadanandam2013colorectal},
ovarian cancer \citep{tothill2008novel},
Parkinson's disease \citep{williams2017parkinson} and
Alzheimer's disease \citep{bredesen2015metabolic}.
Using breast cancer as an example,
the landmark paper by 
\cite{perou2000molecular} was among the first to identify five clinically meaningful subtypes 
(i.e., Luminal A, Luminal B, Her2-enriched, Basal-like and Normal-like) via 
transcriptomic profiles,
and these subtypes had shown distinct disease mechanisms, treatment responses and survival outcomes \citep{van2002gene}.
These molecular subtypes were further followed up by many clinical trial studies for the development of precision medicine \citep{von2012definition, prat2015clinical}.

In the literature, 
many existing clustering algorithms have been successfully applied for disease subtyping,
including hierarchical clustering \citep{eisen1998cluster}, 
K-means \citep{dudoit2002prediction},
mixture model-based approaches, 
\citep{xie2008penalized, mclachlan2002mixture}
and Bayesian non-parametric approaches \citep{qin2006clustering}. 
Due to the high dimensional nature of the transcriptomic data (i.e., large number of genes and relatively small number of samples),
it is generally assumed that only a small subset of intrinsic genes was relevant to disease subtypes \citep{parker2009supervised}.
To capture these intrinsic genes and eliminate noise genes,
sparse clustering methods were proposed, including
sparse K-means or hierarchical clustering \citep{witten2010framework} methods, 
and penalized mixture models \citep{pan2007penalized, wang2008variable}. 
In addition, due to the accumulation of multi-omics dataset (e.g., RNA-Seq, DNA methylation, copy number variation) in public data repositories,
other methods \citep{shen2009integrative, wang2014similarity, huo2017integrative,  cunningham2020particlemdi} sought to identify disease subtypes via integrating multi-omics data.

While promising, 
it has been pointed out that the same omics data may be comprised of multi-facet clusters \citep{gaynor2017identification, nowak2008complementary}.
In other words, 
different configurations of sample clusters defined by separated sets of genes may co-exist.
For example, the desired subtype-related configuration could be driven by disease-related genes, while other configurations of clusters could be driven by sex-related genes, or other genes of unknown confounders. 
Without specifying disease-related genes,
a sparse clustering algorithm tends to identify the configuration with the most distinguishable subtype patterns and separable genes, which optimizes the objective function of the algorithm.
However, it is very likely that the resulting subtypes are not related to the disease, and the selected genes are not relevant to the desired biological process. Nowadays, biomedical studies usually collected comprehensive clinical information, 
which were quite relevant to the disease of interest and could potentially provide guidance for disease subtyping.
Using breast cancer as an example, 
estrogen receptors (ER), progesterone receptors (PR), and human epidermal growth factor receptor 2 (HER2) are hallmarks of breast cancer,
which could provide mechanistic insight about the disease pathology.
Incorporating such clinical information will greatly facilitate the identification of disease related subtypes and intrinsic genes, and enhance the interpretability of disease subtyping results.
Therefore, 
it is urgent to develop a clustering algorithm to fully integrate phenotypic data with high-dimensional omics data. 
Hence, our goal in this paper is to develop a clinical-outcome-guided clustering algorithm. 
Throughout this paper, we use clinical outcome variable and phenotypic data interchangeably, to indicate the potential outcome guidance for the clustering algorithm. 

Many statistical challenges exist to achieve this goal.
(i) Only a subset of intrinsic genes is biological relevant to the disease, 
how to select genes that can best separate disease subtype clusters out of tens of thousands genes from the high dimensional data?
(ii) The clinical outcome variables can be of any data type, 
including continuous, binary, ordinal, count, survival data, etc.
How to incorporate various types of clinical outcome variables,  as a guidance in determining disease subtype clusters?
(iii) Since both the phenotypic data and the intrinsic omics data are utilized, how to balance the contribution of them in determining the subtype clusters?
(iv) How to benchmark whether the resulting subtype clusters are relevant to guidance of the clinical outcome variable?
There have been some efforts in the literature to address parts of these issues. 
\cite{bair2004semi} proposed a two-step clustering method,
in which they pre-selected a list of genes based on the association strength (e.g., Cox score) with the survival outcome, 
and then used the selected genes to perform regular K-means.
However, these pre-selected genes were purely driven by clinical outcomes, and thus might not guarantee to be good separators of the subtype clusters.
Therefore, there is still lack of a unified outcome-guided clustering algorithm that can simultaneously solve all these challenges.

In this paper, we extended from the sparse K-means algorithm by \cite{witten2010framework},
and proposed an outcome-guided clustering (GuidedSparseKmeans) framework by incorporating clinical outcome variable, which will simultaneously solve all these challenges in a unified formulation.
We demonstrated the superior performance of our method in both simulations and gene expression applications of both breast cancer and Alzheimer's disease.

\section{Motivating example}

To demonstrate the motivation of the outcome-guided clustering (GuidedSparseKmeans) method, 
we used the METABRIC \citep{curtis2012genomic} breast cancer gene expression data as an example,
which contained 12,180 genes and 1,870 samples after preprocessing.
Detailed descriptions of the METABRIC data and the preprocessing procedure are available in Section~\ref{sec:metabric}. 
We used the overall survival information as the clinical guidance for the GuidedSparseKmeans (namely, Survival-GuidedSparseKmeans), 
and compared with the non-outcome-guided method (the sparse K-means) and the two-step method (pre-select genes via Cox score, and then perform regular K-means).
We used Silhouette score \citep{rousseeuw1987silhouettes} to benchmark the distinction of subtype patterns of each method, 
for which a larger value represented both better separation between clusters and better cohesion within respective clusters.
We further evaluated the association between the resulting subtypes and the overall survival to examine the biological interpretation of the subtypes derived from each method.

\begin{figure}[h!]
\centering
  \subfloat[Heatmap from the non-outcome-guided method]{\label{fig:1(a)}
    \includegraphics[width=.32\textwidth]{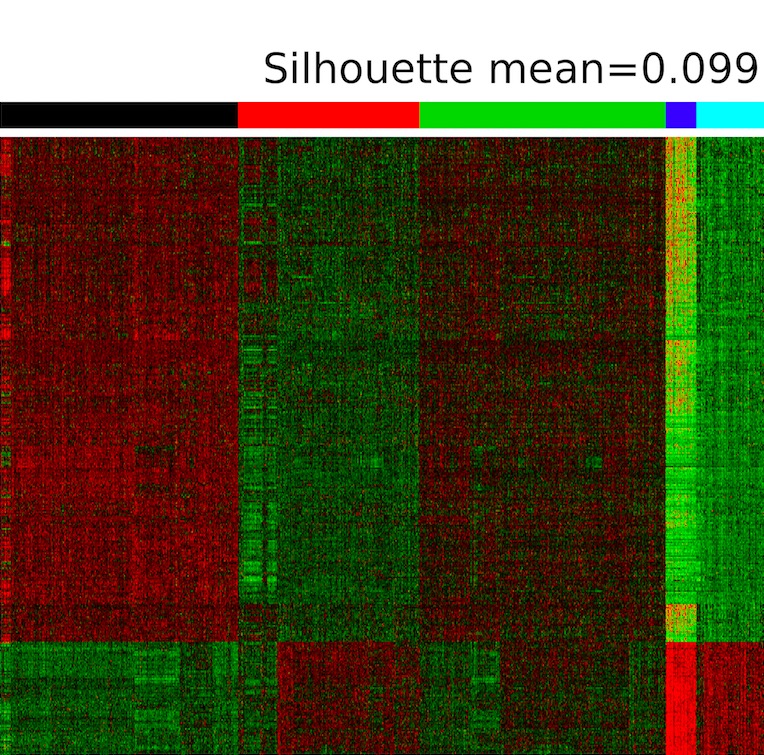}}
~
  \subfloat[Heatmap from the Survival-GuidedSparseKmeans]{\label{fig:1(b)}
    \includegraphics[width=.32\textwidth]{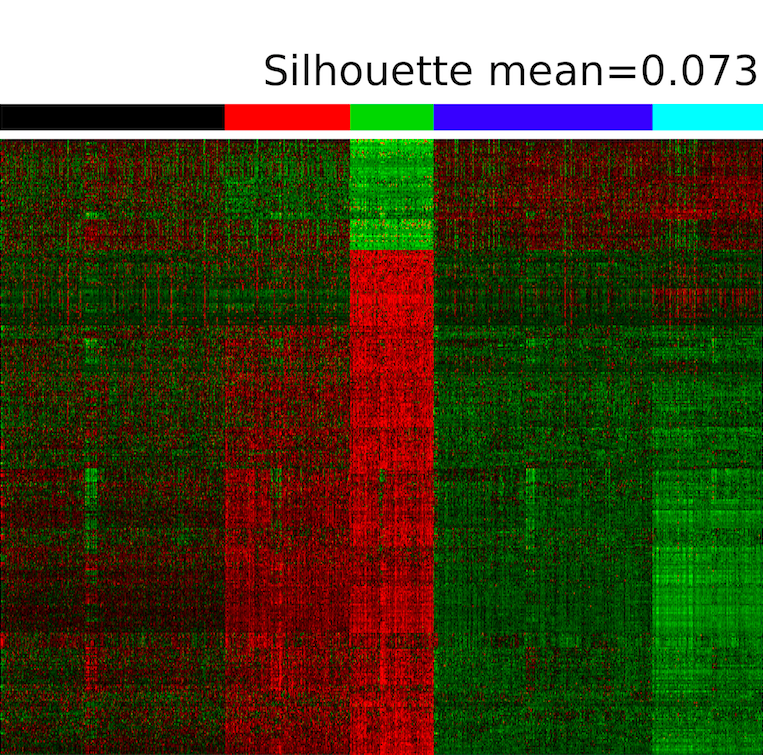}}
~
  \subfloat[Heatmap from the two-step method]{\label{fig:1(c)}
    \includegraphics[width=.32\textwidth]{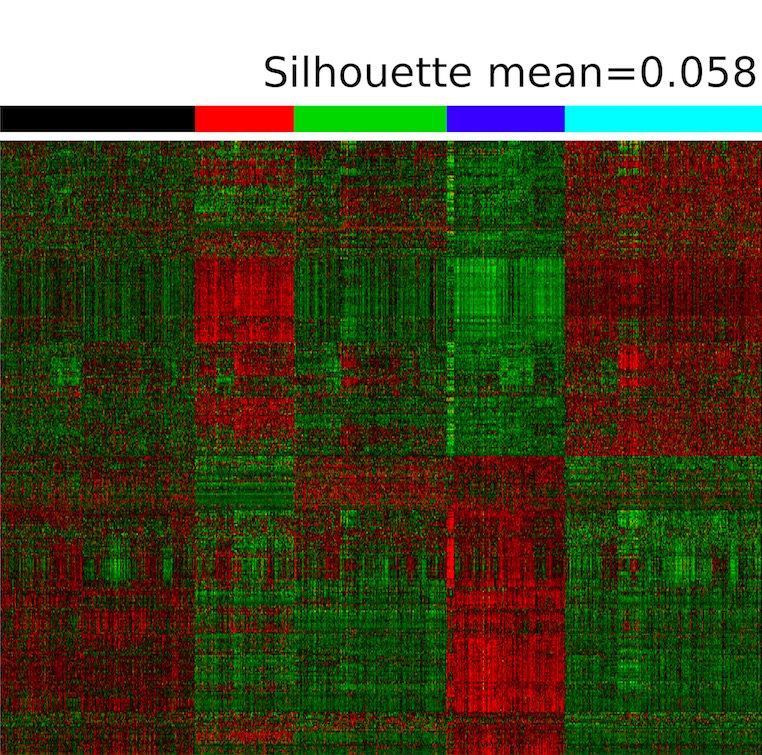}}
    
  \subfloat[Survival curves from the non-outcome-guided method]{\label{fig:1(d)}
    \includegraphics[width=.32\textwidth]{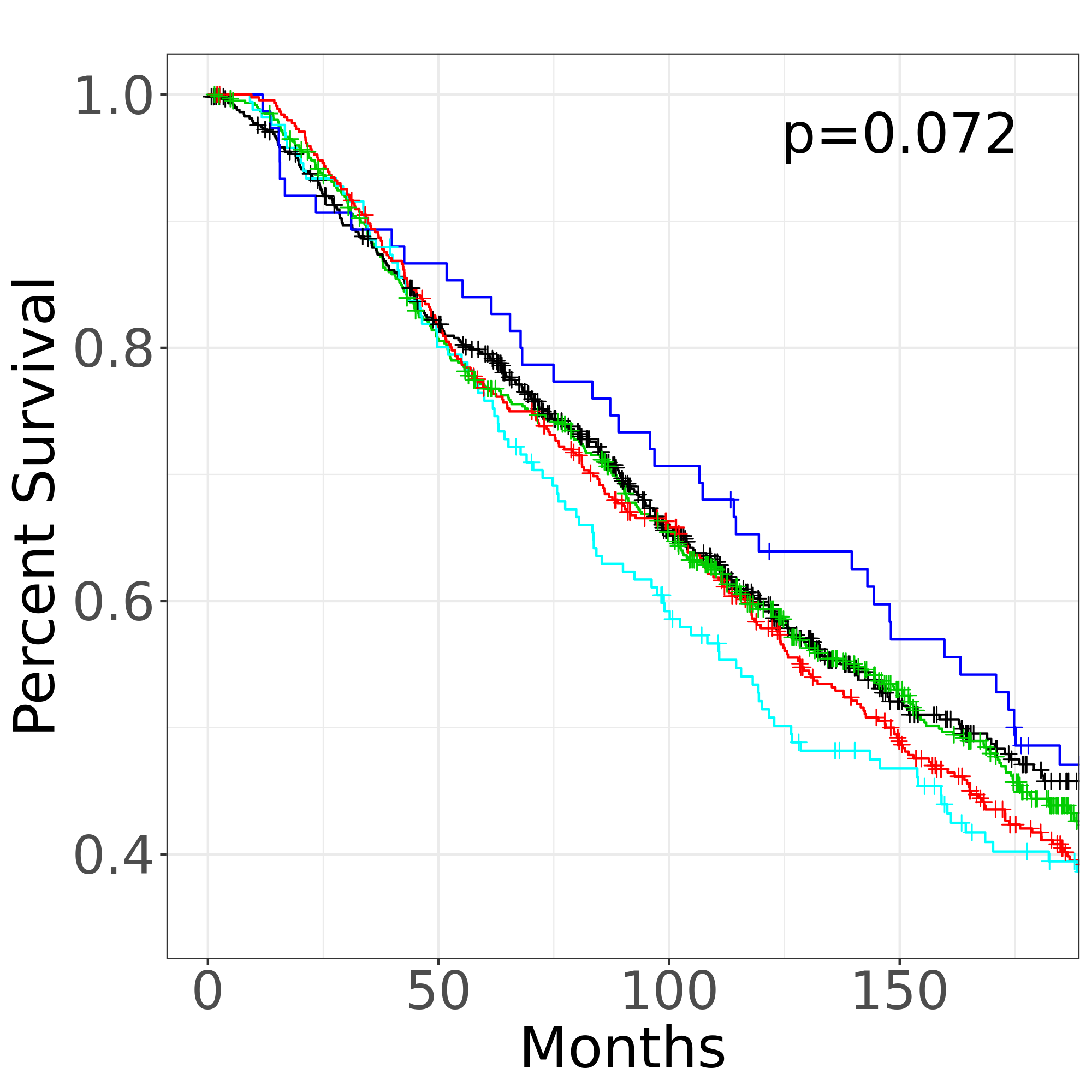}}
~
  \subfloat[Survival curves from the Survival-GuidedSparseKmeans]{\label{fig:1(e)}
    \includegraphics[width=.32\textwidth]{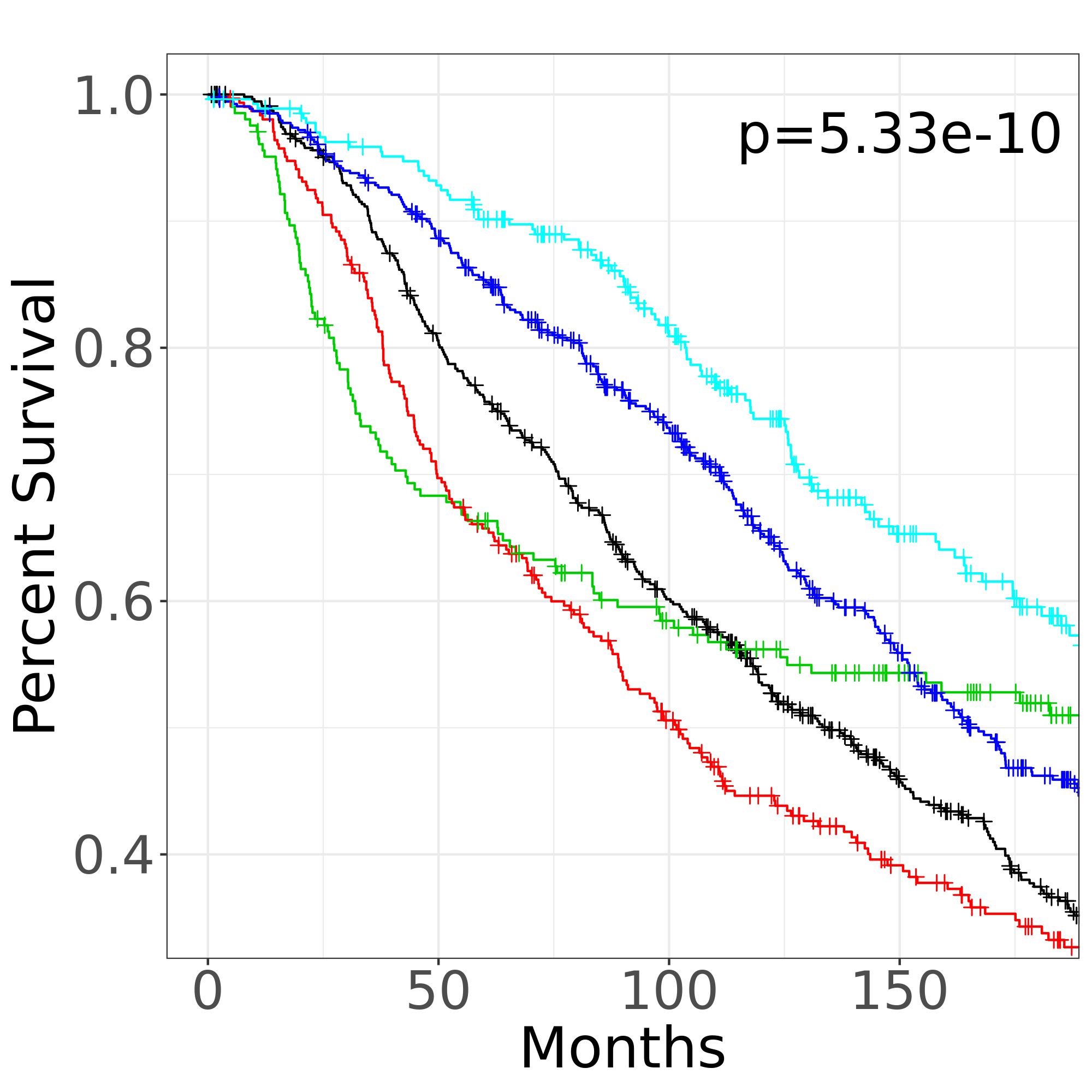}}
~
  \subfloat[Survival curves from the two-step method]{\label{fig:1(f)}
    \includegraphics[width=.32\textwidth]{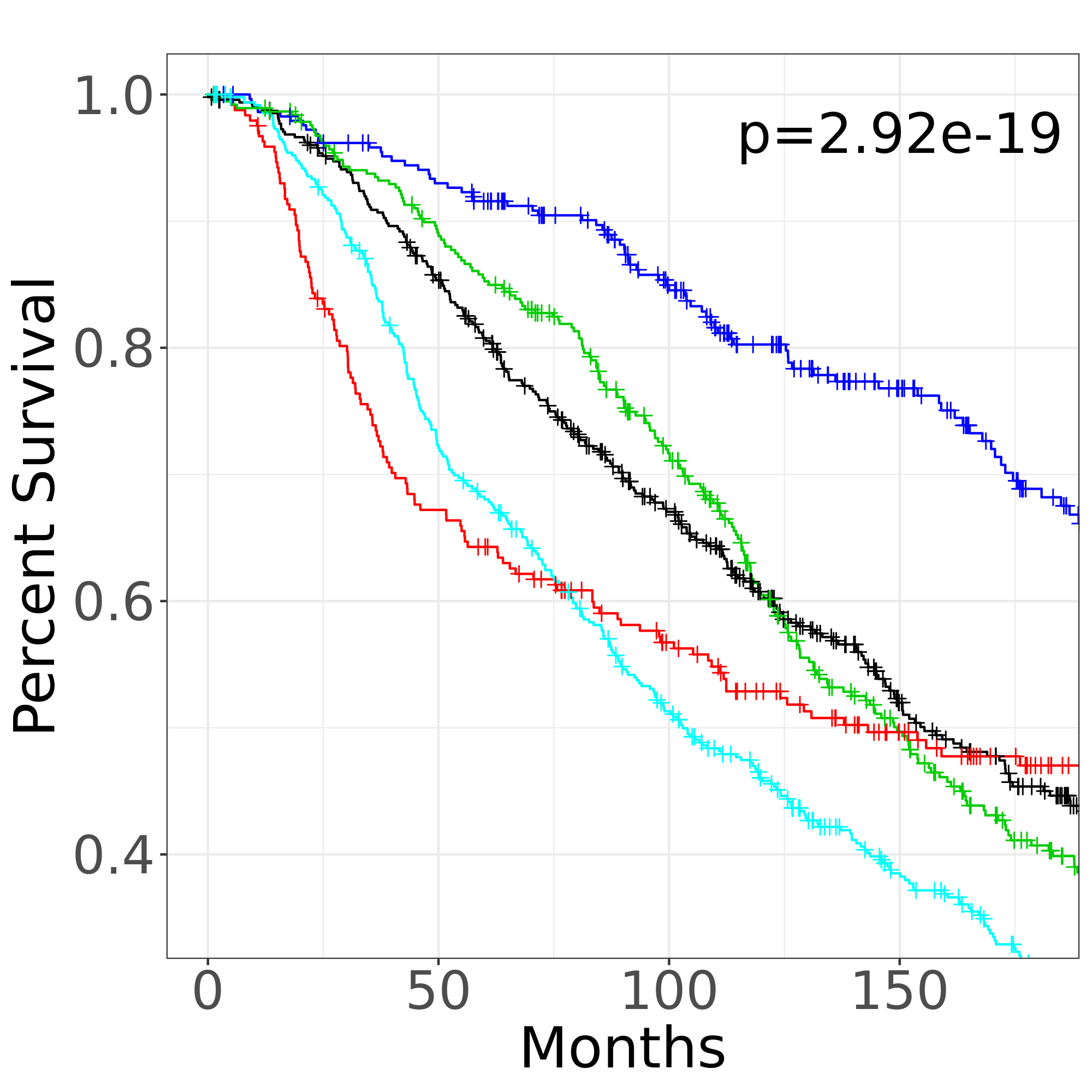}}

\caption{Gene expression heatmap and Kaplan-Meier survival curve of the METABRIC dataset using the non-outcome-guided method (sparse K-means), the Survival-GuidedSparseKmeans, and the two-step method (pre-select genes via Cox score, and then perform regular K-means). In heatmap (a)(b)(c), rows represent genes and columns represent samples. Red color represents higher expression and green color represents lower expression. The samples are divided into 5 clusters with 5 colors in the bar above the heatmaps. ``Silhouette mean'' is the mean Silhouette score of the selected genes, with larger value indicating better separation between clusters. In survival curves (d)(e)(f), the color of the survival curve for each subtype corresponds to the subtype color in the heatmap. The p-values for the survival difference is marked on the top-right corner. Note that these significant p-values are expected as the survival is partly correlated with the clinical variable used.}
\label{fig:1}
\end{figure}

Figure~\ref{fig:1(a)} shows the heatmap of the subtype clustering results from the non-outcome-guided method (the sparse K-means). 
408 genes (on rows) were selected and the 1,870 samples were partitioned into 5 clusters (on columns).
Since the sparse K-means algorithm purely sought for a good clustering performance,
it achieved the most homogeneous heatmap patterns within each subtype cluster (mean Silhouette score = 0.099).
However, the resulting Kaplan-Meier survival curves (Figure~\ref{fig:1(b)}) of the five disease subtypes were not well-separated ($ p = 0.072 $), 
which implied that the subtype clustering results purely driven by the transcriptomic profile may not necessarily be clinically meaningful. 

Figure~\ref{fig:1(c)} shows the heatmap of the subtype clustering results from the Survival-GuidedSparseKmeans. 
384 genes (on rows) were selected and the 1,870 samples were partitioned into 5 clusters (on columns).
The heatmap patterns within each subtype cluster were still visually homogeneous (mean Silhouette score = 0.073).
However, the GuidedSparseKmeans achieved distinct survival separation of the 5 subtype clusters ($p = 5.33 \times 10^{-10}$, Figure~\ref{fig:1(d)}),
which was statistically significant and clinically meaningful.
This is expected because the GuidedSparseKmeans integrated transcriptomic data with phenotypic data, 
seeking for good performance in both clustering result and clinical relevancy. 

Figure~\ref{fig:1(e)} shows the heatmap of the subtype clustering results from the two-step clustering method.
400 genes (on rows) were selected and the 1,870 samples were partitioned into 5 clusters (on columns).
Although the two-step method achieved more distinct survival separation of the 5 subtype clusters ($p = 2.92 \times 10^{-19}$, Figure~\ref{fig:1(f)}),
we observed the large inter-individual variability within each subtype cluster (mean Silhouette score = 0.058), which may further hamper the accurate diagnosis of the future patients.

Collectively,
the sparse K-means algorithm assured homogeneous subtype clustering patterns, but the resulting subtypes might not be clinically meaningful.
The two-step method encouraged the subtypes to be clinically relevant, but might result in high inter-individual variabilities within each subtype cluster.
And our proposed GuidedSparseKmeans simultaneously sought for homogeneous subtype clustering patterns and relevant clinical interpretations, which will be more applicable in biomedical applications.

\section{Method}

\subsection{K-means and sparse K-means}

\subsubsection{K-means algorithm}

Consider $X_{gi}$ the gene expression value of gene $g(1 \le g \le G)$ and subject $i(1 \le i \le n)$. 
Throughout this manuscript, 
we used the gene expression data as the example to illustrate our method, which is easy to be generalized to other types of omics data (e.g., DNA methylation, copy number variation, etc), or even non-omics data.
The K-means method \citep{macqueen1967some} aims to partition $n$ subjects into $K$ clusters such that the within-cluster sum of squares (WCSS) is minimized: 

\begin{equation}
\min_C \sum_{g=1}^{G} WCSS_g(C) = \min_C \sum_{g=1}^{G} \sum_{k=1}^{K} \frac{1}{n_k} \sum_{i,j \in C_k} d_{ij,g},
\label{eq:km}
\end{equation}
where $C_k$ is a collection of indices of the subjects in cluster $k(1\le k \le K)$ with $C=(C_1, C_2, ..., C_K)$; 
$n_k$ is the number of subjects in cluster $k$; 
and $d_{ij,g} = (X_{gi}-X_{gj})^2$ denotes the dissimilarity of gene $g$ between subject $i$ and subject $j$.

\subsubsection{Sparse K-means algorithm}

Since transcriptomic data are usually high-dimensional, with large number of genes $G$ and relatively small number of samples $n$ ($G \gg n$),
it is generally assumed that only a small subset of intrinsic genes will contribute to the clustering result \citep{parker2009supervised}.
To achieve gene selections in the high-dimensional setting, 
\cite{witten2010framework} developed a sparse K-means method with lasso regularization on gene-specific weights.
Lasso regularization is a technique to perform feature selection developed in regression setting \citep{tibshirani1996regression}, which has been used in clustering setting as well \citep{witten2010framework}.
Since directly adding lasso regularization to Equation~\ref{eq:km} would lead to trivial solution (i.e., all weights are 0),
they instead proposed to maximize the weighted $BCSS$ (between cluster sum of square).
This was because minimizing the $WCSS$ of a gene was equivalent to maximizing the $BCSS$ of the gene,
as $BCSS_g(C) = TSS_g - WCSS_g(C)$ and $TSS$ (total sum of square) is a constant for any gene.
Thus the objective function of the sparse K-means algorithm is the following (Equation~\ref{eq:skm}),

\begin{equation}
\max_{C,\textbf{w}} \sum_{g=1}^{G} w_g BCSS_g(C) = \max_{C,\textbf{w}} \sum_{g=1}^{G} w_g \Biggl [\frac{1}{n} \sum_{i,j \in C  } d_{ij,g} -  \sum_{k=1}^{K} \frac{1}{n_k} \sum_{i,j \in C_k} d_{ij,g} \Biggr]
\label{eq:skm}
\end{equation}
subject to $\Vert \textbf{w} \Vert _2 \le 1, \Vert \textbf{w} \Vert_1 \le s, w_g \ge 0\ \forall g$.

In Equation~\ref{eq:skm}, 
$w_g$ denotes the weight for gene $g$ with $\mathbf{w}=(w_1, w_2,...,w_G)$; 
the $l_1$ norm penalty $\| \textbf{w} \|_1$ will facilitate gene selection (i.e., only a small subset of genes have non-zero weights);
$s$ is a tuning parameter to control the number of nonzero weights (larger $s$ will result in more selected genes);
and the $l_2$ norm penalty $\| \textbf{w} \|_2$ will encourage selecting more than one genes
because otherwise, only one gene with the largest $BCSS$ will be selected.

\subsection{GuidedSparseKmeans}

In order to incorporate the contribution from a clinical outcome variable, 
we proposed to use a gene specific award term $U_g$.  
$U_g$ represents the association strength between gene $g$ and a clinical outcome variable, which will help guide gene selections.
For example, $U_g = U(\textbf{x}_g, \textbf{y})$, 
where $\textbf{x}_g = (X_{g1}, \ldots, X_{gn})^\top \in \mathbb{R}^n$ is the expression levels of gene $g$, and 
$\textbf{y} = (y_1, \ldots, y_n)^\top \in \mathbb{R}^n$ is the vector of clinical outcome.
The objective function of the GuidedSparseKmeans is as follows:

\begin{equation}
\max_{C,\textbf{w}} \sum_{g=1}^{G} w_g \Biggl[\frac{BCSS_g(C)}{TSS_g} + \lambda U_g \Biggr]
\label{eq:gskm}
\end{equation}
subject to $\| \textbf{w} \|_2 \le 1, \Vert \textbf{w} \Vert_1 \le s, w_g \ge 0\ \forall g$,\\
where $\lambda$ is the tuning parameter controlling the balance between the clustering separation ability of gene $g$ (i.e., $\frac{BCSS_g(C)}{TSS_g}$) and the guidance of the clinical outcome variable $U_g$. 
$BCSS_g$ is normalized against $TSS_g$ such that the range of $\frac{BCSS_g(C)}{TSS_g}$ is $[0,1]$, which is comparable to the range of $U_g$.
Thus, by maximizing Equation~\ref{eq:gskm},
we will select genes with both strong clustering separation ability and strong association with the clinical outcome variable.

In order to accommodate various types of clinical outcome variables, 
we proposed to calculate $U_g$ based on a univariate regression model $f_g$,
where a clinical outcome variable is the dependent variable and the expression of gene $g$ is the independent variable.
To be specific, 
the guiding term $U_g$ is defined as the pseudo R-squared proposed by \cite{cox1989analysis} for $f_g$:
$$U_g =1 - \Bigl[ \frac{L(f_0)}{L(f_g)} \Bigr] ^{2/n} $$
where $n$ is the number of subjects; $L(f_0)$ is the likelihood of null model (i.e., intercept only model); and $L(f_g)$ is the likelihood of the model $f_g$. 
Since the univariate regression model $f_g$ can be linear models, generalized linear models, Cox models, or other regression models,
the proposed $U_g$ can accommodate a wide variety of clinical outcome variables (e.g., continuous, binary, ordinal, count, survival data, etc). 
We acknowledge that $U_g$ could be also defined using the pseudo R-squared proposed by \cite{mcfadden1977quantitative}, \cite{tjur2009coefficients}, \cite{nagelkerke1991note}, \cite{efron1978regression} or \cite{mckelvey1975statistical}, or simply the absolute value of the Pearson correlation coefficient between $\textbf{x}_g$ and $\textbf{y}$. 

To quantify whether the subtype clustering results are relevant to the clinical outcome variable,  
we further proposed a relevancy score by calculating the Pearson correlation of $\textbf{w}^*$ and $\textbf{u}^*$, 

\begin{equation}
\mbox{Rel} = \mbox{cor}(\textbf{w}^*, \textbf{u}^*)
\label{eq:relevancy}
\end{equation}
where $\textbf{w}^* = \{w_g: w_g \ne 0\}$ and  $\textbf{u}^* = \{U_g: w_g \ne 0\}$. 
Since $\textbf{w}^*$ represents the non-zero weights of the selected genes that are associated with the clustering result, 
and $\textbf{u}^*$ is associated with the clinical outcome variable,
the relevancy score indirectly quantifies the association between the subtype clustering results and the guidance of the clinical outcome variable.

\begin{remark}
In Equation~\ref{eq:gskm},
$\lambda$ controls the balance between the separation ability of gene $g$ (i.e., $\frac{BCSS_g(C)}{TSS_g}$) and the guidance of the clinical outcome variable $U_g$. 
When $\lambda = 0$, Equation~\ref{eq:gskm} reduces to Equation~\ref{eq:skm}.
Therefore, the sparse $K$-means algorithm is a special case of the GuidedSparseKmeans when $\lambda = 0$. 
When $\lambda \rightarrow \infty$, 
the gene-specific weights are purely driven by the clinical outcome variable (i.e., only genes with large $U_g$ will be selected). 
Equation~\ref{eq:gskm} reduces to the two-step clustering algorithm,
and the clustering results are determined by the weighted K-means algorithm based on these pre-selected genes.
Therefore, the two-step algorithm is also essentially a special case of the GuidedSparseKmeans when $\lambda \rightarrow \infty$. 
And in Section \ref{sec:tuning} we proposed to use sensitivity analysis to find the optimal $\lambda$ to balance these two terms.
\end{remark}

\subsection{Optimization}
\label{sec:opt}

Given the selected values of $K$, $\lambda$, and $s$, the optimization procedure is carried out as the following:

\begin{enumerate}[1.]
	\item Denote $U_g^*=U_g\times \mathbb{I}\{U_g \ge U_{[r]}\}$, where $\mathbb{I}$ is an indicator function and $U_{[r]}$ is the $r^{th}$ largest element in vector $\textbf{u} = (U_1, \ldots, U_g)^\top$. 
    We chose $r=400$ throughout this manuscript. 
    In the later breast cancer application example, we performed sensitivity analysis to demonstrate that the difference choice of $r$ did not have a large impact on the clustering results.
    Initialize $\mathbf{w}$ as $w_g=\frac{U_g^*}{\sum_{g=1}^{G} U_g^*} \times s$ \label{step:ini}.
	\item Holding $\mathbf{w}$ fixed, update $C_k(1\le k \le K)$ by weighted K-means: \label{step:updateCs}
	$$C(\mathbf{w}) = \argmax_{C} \sum_{g=1}^{G} w_g BCSS_g(C) $$
	20 random initial centers were considered in the starting of the weighted K-means, and the cluster result with the maximized objective function (among these 20 trials) was adopted as the output for the weighted K-means.
	\item Holding $ C=(C_1, C_2, ..., C_K)$ fixed, update $\mathbf{w}$ as the following:
    $$w_g (C) =\frac{S(a_g, b)}{\| S(a_g, b) \| _2},$$
    where $S$ is the soft-thresholding operator defined as $S(x,c) = {\rm sign}(x) \max(x-c, 0)$; $a_g = \frac{BCSS_g(C)}{TSS_g} + \lambda U_g$; $b > 0$ is chosen such that $\| \mathbf{w} \| _1 =s$, otherwise, $b=0$ if  $\Vert \mathbf{w} \Vert  _1 < s$.
    \label{step:updateWs}
	\item Iterate Step 2-3 until the stopping rule 
	$\frac{\| \mathbf{w}^t - \mathbf{w}^{t-1} \| _1}{\| \mathbf{w}^{t-1} \| _1} < 10^{-4}$ is satisfied, where {\color{blue} t} is the number of iterations.
\end{enumerate}

In Step~\ref{step:updateCs}, the optimization of clusters $C$ is essentially applying a weighted K-means algorithm on the original gene expression data, 
since the weights will change the relative scale of the data, 
and the $TSS_g$ and $\lambda U_g$ are not functions of $C$.
In Step~\ref{step:updateWs}, fixing the clusters $C$, the optimization of weights $\mathbf{w}$ is a convex problem, and a closed form solution of $\mathbf{w}$ can be derived by Karush-Kuhn-Tucker (KKT) conditions \citep{boyd2004convex, witten2010framework}.

\subsection{Selection of tuning parameter} 
\label{sec:tuning}

The GuidedSparseKmeans objective function (Equation~\ref{eq:gskm}) has three tuning parameters: 
$K$ is the number of clusters; 
$s$ controls the number of selected genes;
and $\lambda$ controls the balance between the clustering separation ability and the guidance from the clinical outcome.
Below are our guidelines on how to select these tuning parameters.

\subsubsection{Selection of $K$} 
\label{sec:tuningK}

How to select number of clusters $K$ has been widely discussed in the literature \citep{milligan1985examination, kaufman2009finding, sugar2003finding, tibshirani2005cluster, tibshirani2001estimating}.
Here, we suggested $K$ to be estimated using the gap statistics \citep{tibshirani2001estimating} or prediction strength \citep{tibshirani2005cluster}. 
Alternatively, $K$ can be set as a specific number in concordance with the prior knowledge of the disease.
In our simulation, we selected top $N$ genes (e.g., $N=400$) with the largest $U_g$'s, and used gap statistics to select $K$ based on the $N$ pre-selected genes.
The gap statistics is the difference between observed WCSS and the background expectation of WCSS calculated from random permutation of the original data.
The rationale behind this is that the underlying true number of clusters K should occur when this difference (observed WCSS – background expected) is maximized \citep{tibshirani2001estimating}.

\subsubsection{Selection of $\lambda$} 
\label{sec:tuningLambda}

The tuning parameter $\lambda$ controls the balance between the separation ability of gene $g$ (i.e., $\frac{BCSS_g(C)}{TSS_g}$) and the guidance of the clinical outcome variable $U_g$. 
We recommend to use sensitivity analysis to select $\lambda$. 
When $\lambda \rightarrow \infty$, 
the selected genes as well as the clustering results are purely driven by the guidance of the clinical outcome variable $U_g$.
As $\lambda$ gradually decreases, we expect stable results of gene selection and clustering results, 
until a certain transition point.
After the transition point, we expect the separation ability $\frac{BCSS_g(C)}{TSS_g}$ becomes dominant, and both gene selection and clustering results may change. 
Therefore, our rationale is to identify the transition point $\lambda^*$,
which is the smallest $\lambda$ such that the clustering results and the gene selection results are stable with respect to any $\lambda \ge \lambda^*$. 
Below is the detailed algorithm:

\begin{enumerate}[1.]
	\item 
    Let $\lambda^M = (\lambda_1,...,\lambda_{m},...,\lambda_{M})$, $\lambda_m = 0.25m$ for $1 \le m \le M$. 
    Obtain the clustering result $C(\lambda_m, s, K)$ and gene weight $\textbf{w}(\lambda_m, s, K)$ for each $\lambda_m(1 \le m \le M)$ by applying the GuidedSparseKmeans, where $K$ is pre-estimated, and $s$ is pre-set so that $q$ genes can be selected by the GuidedSparseKmeans ($q$ is the proportion of the total number of genes in the expression dataset and $q$ is recommended to set between 3\% and 6\%). 
    We set $M=10$ throughout this paper.
    Since both $s$ and $K$ are fixed, 
    we use $C(\lambda_m)$ to denote $C(\lambda_m, s, K)$, 
    and $\textbf{w}(\lambda_m)$ to denote $\textbf{w}(\lambda_m, s, K)$. 
    For $1 \le m \le M-1$, 
    we further calculate the adjusted Rand index \citep{hubert1985comparing} (ARI) between $C(\lambda_m)$ and $C(\lambda_{m+1})$ as $A(m, m+1)$. 
    Similarly, we calculate the Jaccard index \citep{jaccard1901etude} between $\mathbb{I}\{\textbf{w}(\lambda_m)=0\}$ and $\mathbb{I}\{\textbf{w}(\lambda_{m+1})=0\}$ as $J(m, m+1)$, where $\mathbb{I}$ is an indicator function. 
    ARI measures the agreement of two clustering results, and the Jaccard index measures the agreement of two gene sets (see Section~\ref{sec:result} for details).

	\item For $2 \le m \le M-2$, 
	calculate 
	$\mu_A(m) = \frac{1}{M-m} \sum_{i = m}^{M-1} A(i, i+1)$;
	$\sigma_A(m)$ as the standard deviation of $A(i,i+1)$'s for $m\le i \le M-1$;
	$\mu_J(m) = \frac{1}{M-m} \sum_{i = m}^{M-1} J(i, i+1)$;
	and $\sigma_J(m)$ as the standard deviation of $J(i,i+1)$'s for $m\le i \le M-1$. 
	
	\item We calculate $m^{A*}$ as the largest $m$ such that $A(m-1,m) < \mu_A(m) - 2\max\{\sigma_A(m), \delta\}$, where we set the fudge parameter $\delta = 0.05$ to avoid the zero variation.
	Similarly, we calculate $m^{J*}$ as the largest $m$ such that $J(m-1,m) < \mu_J(m) - 2\max\{\sigma_J(m), \delta \}$.
	$m^{A*}$ and $m^{J*}$ mimic the transition points for the clustering result and gene selection result, respectively.
    \item We select $\lambda$ as $\lambda^* = \max \{\lambda_{m^{A*}}, \lambda_{m^{J*}}\}$
    \item If $A(m,m+1)$ (or $J(m, m+1)$) is very stable for all $1 \le m \le M - 1$,
	it is possible that $m^{A*}$ (or $m^{J*}$) will not be obtained. Under this scenario, $\lambda$ is not sensitive to clustering (or gene selection result). 
    Since $\lambda^* = \max \{\lambda_{m^{A*}}, \lambda_{m^{J*}}\}$, 
    we will set $m^{A*}=1$ (or $m^{J*}=1$) to let its alternative component $m^{J*}$ (or $m^{A*}$) to decide the proper $\lambda$. 

\end{enumerate}

Based on the above algorithm, 
the clustering results and gene selection results will not be sensitive to $\lambda$ if $\lambda \ge \lambda^*$.

\subsubsection{Selection of $s$} 
\label{sec:tuningS}

After $K$ and $\lambda$ are selected, we extend the gap statistics procedure in the sparse K-means algorithm \citep{witten2010framework} to estimate $s$, which controls the number of selected genes:

\begin{enumerate}[1.]

    \item Pre-specify a set of candidate tuning parameter $s$.

    \item Create $B$ permuted datasets $X^{(1)}, X^{(2)},..., X^{(B)}$ by randomly permuting subjects for each gene feature of the observed data $X$. 

    \item Compute the gap statistics for each candidate tuning parameter $s$ as the following:
    $$Gap(s) = O(s) - \frac{1}{B} \sum_{b=1}^{B} O_b(s)$$
    where $O(s) = \log \bigl(\sum_{g=1}^{G} w_g^* BCSS_g(C^*)\bigr)$ is
    the log weighted $BCSS$ value on the observed data $X$ with $w_g^*$ and $C^*$ being the optimum solution. 
    Similarly, $O_b(s)$ is the log weighted $BCSS$ value on the permutated data $X^{(b)}$ and their corresponding optimum solutions. 

    \item Choose $s^*$ such that the $Gap(s^*)$ is maximized. 
\end{enumerate}

\begin{figure}[h!]
\centering
  \subfloat[Gap statistics for each $K$]{\label{fig:2(a)}
    \includegraphics[width=.45\textwidth]{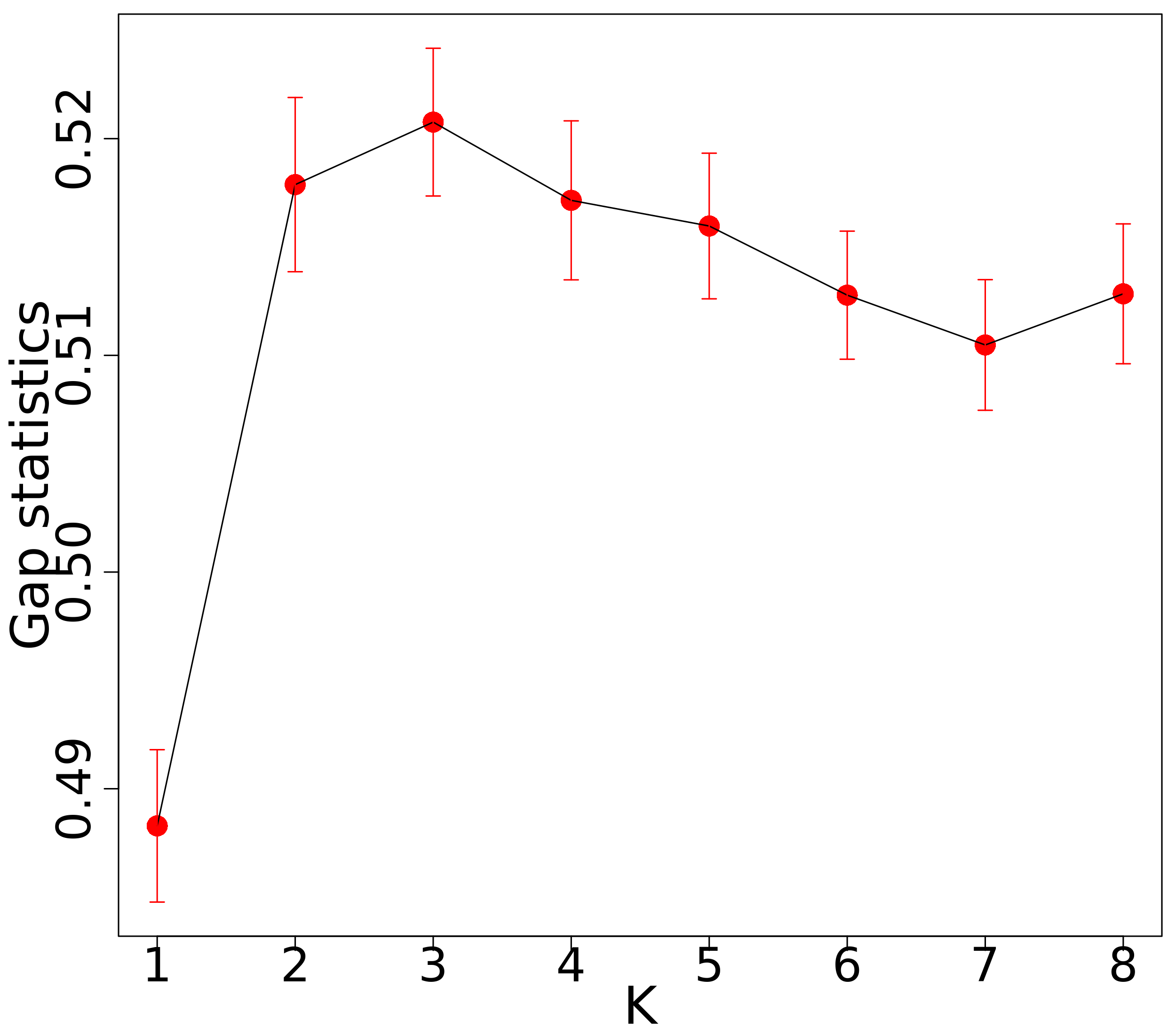}}
~
  \subfloat[Gap statistics for each $s$]{\label{fig:2(b)}
    \includegraphics[width=.45\textwidth]{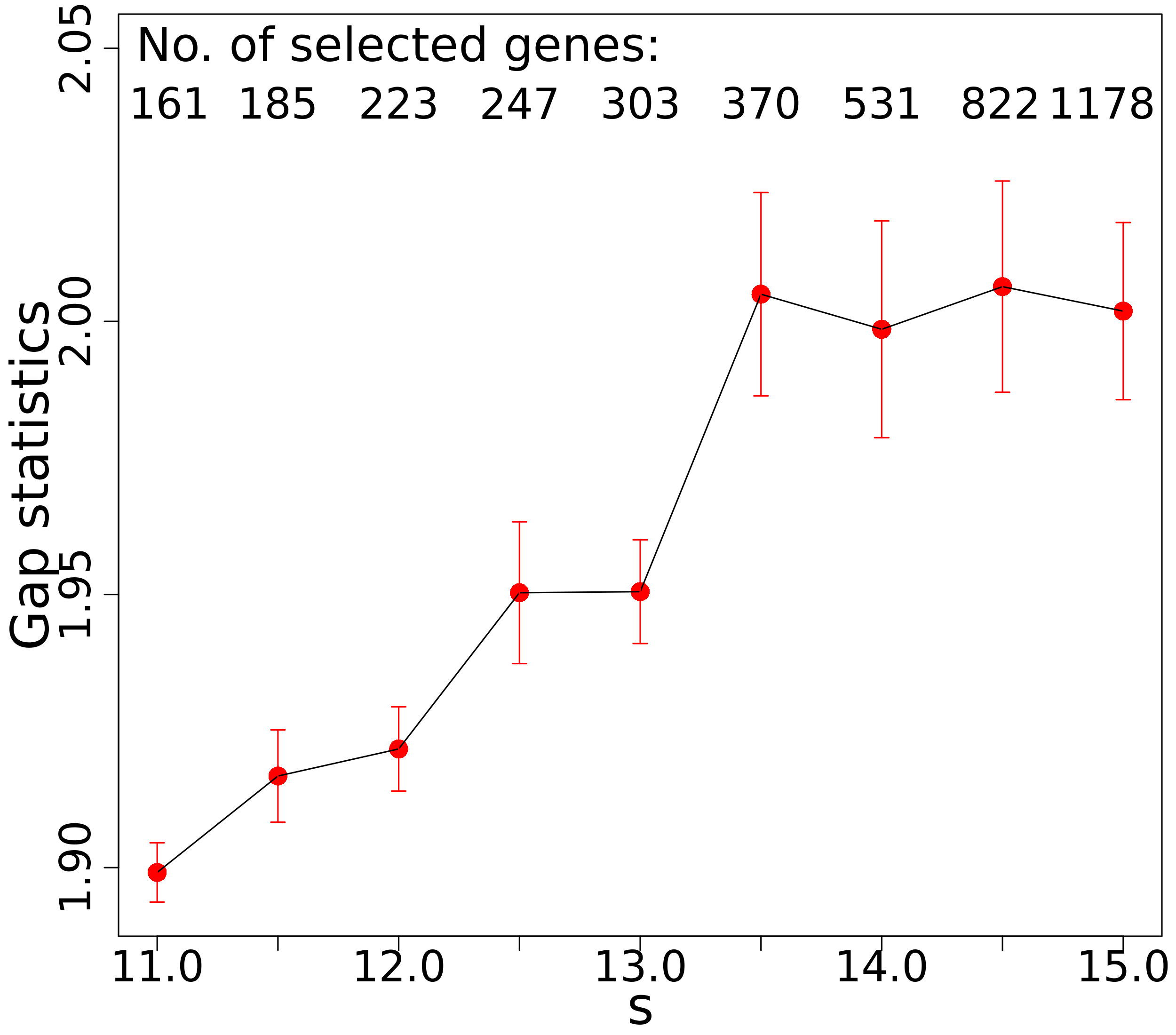}}
    
  \subfloat[ARI values for the comparison between two clustering results obtained from $\lambda_m$ and $\lambda_{m+1}$ $(1 \le m \le 9)$]{\label{fig:2(c)}
    \includegraphics[width=.45\textwidth]{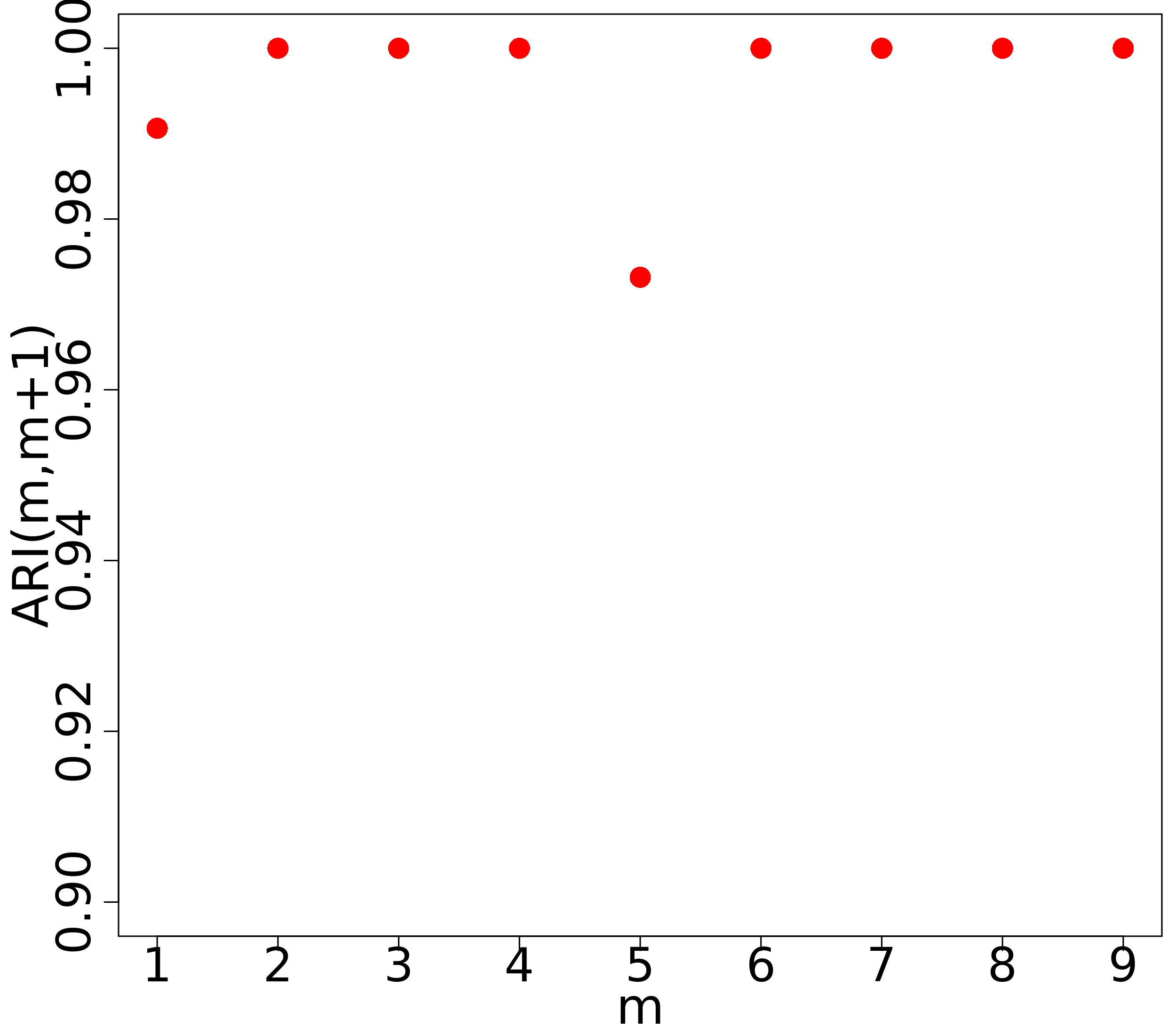}}
~
  \subfloat[Jaccard index values for the comparison between two gene selection results obtained from $\lambda_m$ and $\lambda_{m+1}$ $(1 \le m \le 9)$]{\label{fig:2(d)}
    \includegraphics[width=.45\textwidth]{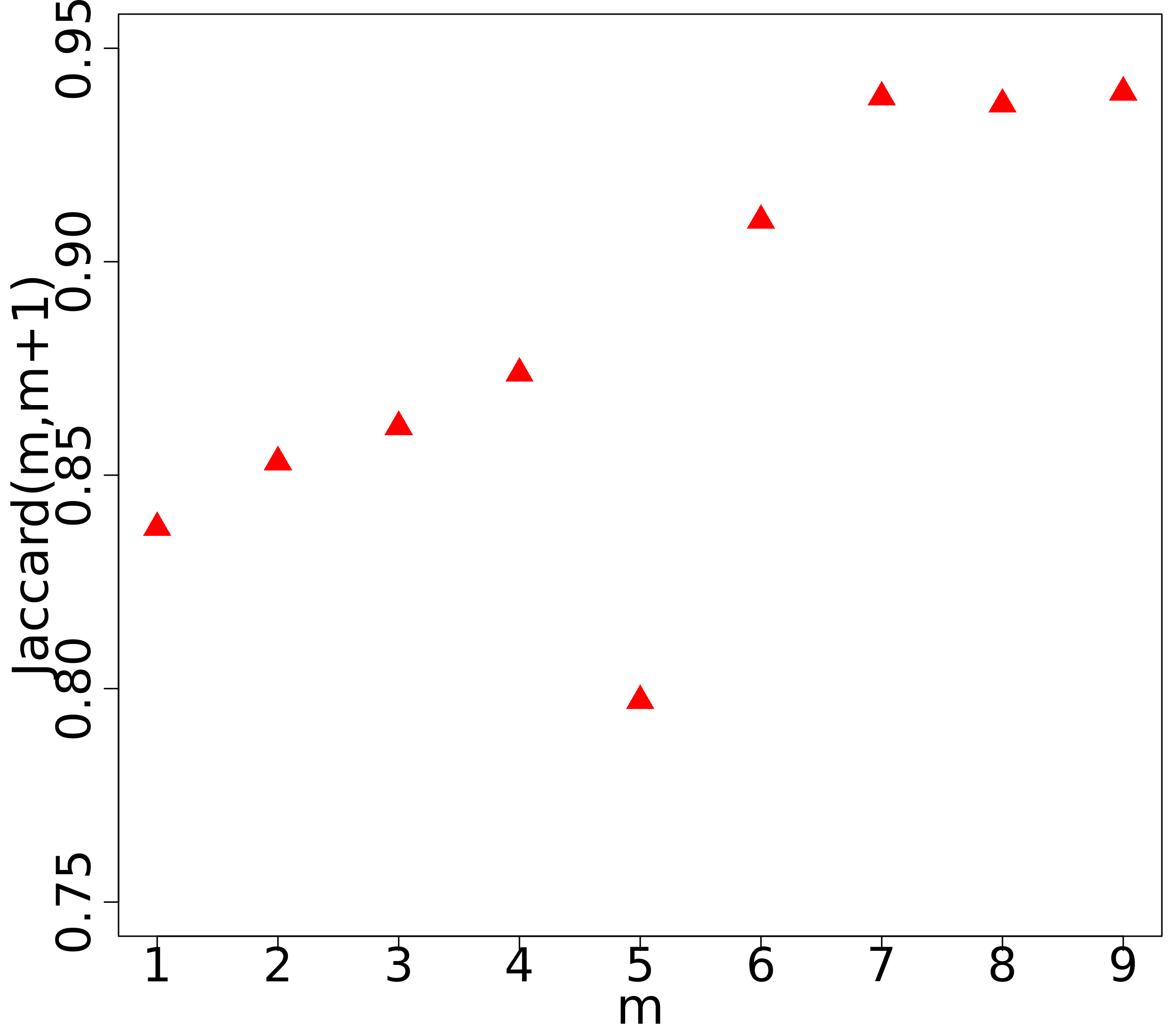}}

\caption{Selection of tuning parameters in a simulated dataset with biological variation $\sigma_1=3$. There are total 326 samples with 104, 110, and 112 samples in each clusters and total 9,988 genes including 389 true intrinsic genes, 1,599 confounding genes, and 8,000 noise genes. Figure 2(a): Gap statistics to select $K$. The gap statistics is maximized at $K=3$. Figure 2(b): Gap statistics to select $s$. The number of selected genes corresponding to each candidate $s$ is showed on the top. The gap statistics is maximized at $s=13.5$, which is corresponding to 370 genes. Figure 2(c)(d): Sensitivity analysis to select $\lambda$. The values of $A(m,m+1)$ and $J(m,m+1)$ are calculated for each $m$. $m^{A*}=1$ and $m^{J*}=6$ are transition points in (c) and (d) respectively. The estimated $\lambda^*$ is 1.5 ($\max \{\lambda_{m^{A*}}, \lambda_{m^{J*}}\}$).}
\label{fig:2}
\end{figure}

\section{Result}

\label{sec:result}

We evaluated the performance of the GuidedSparseKmeans on simulation datasets and two gene expression profiles of breast cancer and Alzheimer’s disease, and compared with the non-outcome-guided method (the sparse K-means) and the two-step clustering method (regular K-means with pre-selected genes via Cox score). 
We used the adjusted Rand index \citep{hubert1985comparing} (ARI) to benchmark the clustering performance, which evaluates the similarity between a clustering result and the underlying truth (ranges from 0 [random agreement] to 1 [perfect agreement]). 
We used the Jaccard index \citep{jaccard1901etude} to benchmark the similarity of the selected genes to the underlying intrinsic genes, which is defined as the ratio of the number of genes from both gene sets to the number of genes from either one of the gene sets (ranges from 0 [no intersection between the two gene sets] to 1 [identical gene sets]).

\subsection{Simulation} 
\label{sec:simu}

\subsubsection{Simulation setting}
\label{sec:simuSetting}

To evaluate the performance of the GuidedSparseKmeans and to compare with the non-outcome-guided method (the sparse K-means), 
we simulated a study with $K=3$ subtypes. 
To best preserve the nature of genomic data, we simulated intrinsic genes (i.e., genes defining subtype clusters), confounding impacted genes (i.e., genes defining other configurations of clusters), and noise genes with correlated gene structures \citep{song2014hypothesis, huo2016meta}. 
In addition, we generated a continuous outcome variable as the clinical guidance. 
Below is the detailed data generative process.

\begin{enumerate}
	\item Intrinsic genes.
    \begin{enumerate}[1.]
	    \item Simulate $N_k \sim POI(100)$ subjects for subtype $k (1 \le k \le K)$ and the number of subjects in the study is $N=\sum_k N_k$.
	    For each simulated data, the expected number of subjects is 300.

	    \item Simulate $M=20$ gene modules ($1 \le m \le M$). 
        Denote $n_m$ as the number of features in module $m$ ($1 \le m \le M$). 
        In each module, sample the number of genes $n_m$ by $n_m \sim POI(20)$. 
        Therefore, there will be an average of 400 intrinsic genes. 
	    
	    \item Denote $\theta_k$ as the baseline gene expression of subtype $k(1 \le k \le K)$ of the study and $\theta_k = 2 + 2k$. 
        Denote $\mu_{km}$ as the template gene expression of subtype $k$ and module $m(1 \le m \le 20)$. 
        We calculate the template gene expression by $\mu_{km} = \alpha_m \theta_k + N(0, \sigma_0^2)$, where $\alpha_m$ is the fold change for each module  $m$  and $\alpha_m \sim UNIF \bigl( (-2, -0.2) \cup (0.2, 2) \bigr)$. $\sigma_0$ is set to be 1.
	    
	    \item  Add biological variation $\sigma_1^2$ to the template gene expression and generate $X_{kmi}^{'} \sim N(\mu_{km}, \sigma_1^2)$ for each module $m(1 \le m \le 20)$, subject $i(1 \le i \le N_k)$ of subtype $k$. 
	    $\sigma_1$ is set to be 3 unless otherwise specified.

	    \item Simulate the covariance matrix $\Sigma_{km}$ for genes in subtype $k$ and module $m(1 \le m \le 20)$. 
        First sample $\Sigma_{km}^{'} \sim W^{-1}(\phi, 60)$, where $\phi = 0.5I_{n_m \times n_m} + 0.5J_{n_m \times n_m}$, $W^{-1}$ denotes the inverse Wishart distribution, $I$ is the identity matrix and $J$ is the matrix with all elements equal to 1. 
        The $\Sigma_{km}$is calculated by standardizing  $\Sigma_{km}^{'}$such that all the diagonal elements are equal to 1. 
	    \item Simulate gene expression values for subject $i$ in subtype $k$ and module $m$ as $(X_{1kmi}, ..., X_{n_mkmi})^\top \sim MVN(X_{kmi}^{'}, \Sigma_{km})$, where $1 \le k \le 3$, $1 \le m \le 20$ and $1 \le i \le N_k$. 
    \end{enumerate}
    \item Phenotypic variables.
    \begin{enumerate}[1.]
	    \item Simulate clinical outcomes as $Y_{ki} \sim N(\theta_k, \sigma_2^2)$ for subject $i$ and in subtype $k$, where $1 \le i \le N_k$ and $1 \le k \le 3$. 
        The clinical outcome variation $\sigma_2$ is set to be 8 such that the magnitude of $U_g$ of the intrinsic genes in this simulation is similar to the breast cancer example (Section~\ref{sec:metabric}).
	\end{enumerate}
    \item Confounding impacted genes.
    \begin{enumerate}[1.]
	    \item Simulate $V=4$ confounding variables. 
        Confounding variables can be demographic factors such as race, gender, or other unknown confounding variables. 
        These variables will define other configurations of sample clustering patterns, and thus complicate disease subtype discovery.
        For each confounding variable $v(1 \le v \le V)$, we simulate $R=20$ modules. 
        For each module $r_v(1 \le r_v \le R)$, sample number of genes $n_{r_v} \sim POI(20)$. 
        Therefore, there will be an average of 1,600 confounding impacted genes.
        
        \item For each confounding variable $v$ in the study, the $N$ samples are randomly divided into $K$ subclasses, which represents the configuration of clusters for confounding variables.
        
        \item Denote $\mu_{kr_v}$ as the template gene expression in subclass $k(1 \le k \le K)$ and module $r_v(1 \le r_v \le 20)$. 
        We calculate the template gene expression by $\mu_{kr_v} = \alpha_{r_v} \theta_k + N(0, \sigma_0^2)$, where $\alpha_{r_v}$ is the fold change for each module $r_v$ and $\alpha_{r_v} \sim UNIF \bigl( (-2, -0.2) \cup (0.2, 2) \bigr)$, $\theta_k$ is the same as the one in a3. 
        
        \item Add biological variation $\sigma_1^2$ to the template gene expression and generate $X_{kr_vi}^{'} \sim N(\mu_{kr_v}, \sigma_1^2)$. 
        Similar to Step a5 and a6, we simulate gene correlation structure within modules of confounder impacted genes.
	\end{enumerate}
    \item Noise genes.
    \begin{enumerate}[1.]
	    \item Simulate 8,000 noninformative noise genes denoted by $g(1\le g \le 8,000)$. 
        We generate template gene expression $\mu_g \sim UNIF(4, 8)$. 
        Then, we add noise $\sigma_3 = 1$ to the template gene expression and generate $X_{gi} \sim N(\mu_g, \sigma_3^2)$.
	\end{enumerate}
\end{enumerate}

For each simulated data, the expected number of genes is 10,000, including 400 intrinsic genes, 1,600 confounding impact genes, and 8,000 noise genes.

\subsubsection{Simulation results}

In this section, we first showed the tuning parameter estimation results of the GuidedSparseKmeans using one simulation dataset. 
As shown in Figure~\ref{fig:2(a)}, the gap statistics introduced in Section~\ref{sec:tuningK} successfully identified $K=3$ as the optimal number of clusters, which had the largest gap statistics.
The guidance term $U_g$ was calculated as the $R^2$ from univariate linear regressions.
We applied the sensitivity analysis algorithm introduced in Section~\ref{sec:tuningLambda} to select $\lambda$ for the simulated dataset.
Figure~\ref{fig:2(c)} shows the value of $A(m,m+1)$ for each $m (1 \le m \le 9)$ and Figure~\ref{fig:2(d)} shows the value of $J(m, m+1)$ for each $m (1 \le m \le 9)$.
Then, $\lambda_{m^{A*}}=0.25$ $(m^{A*}=1)$ and $\lambda_{m^{J*}}=1.5$ $(m^{J*}=6)$ were obtained.
Therefore, we set $\lambda^*=1.5$ in this specific simulation via sensitivity analysis.
We applied the gap statistics algorithm introduced in Section~\ref{sec:tuningS} to select $s$ for the simulated dataset.
Figure~\ref{fig:2(b)} shows the gap statistics for each candidate $s$ ($s = 11, 11.5, \ldots, 15$). 
The tuning parameter $s^*$ was chosen to be 13.5 and the number of selected genes was 370, which was close to the underlying truth (389 intrinsic genes).

Then, we compared the performance of the GuidedSparseKmeans and the non-outcome-guided algorithm (the sparse K-means).
The number of clusters was estimated via the gap statistics in Section~\ref{sec:tuningK}.
The tuning parameter $\lambda$ for the GuidedSparseKmeans for each single simulation was selected based on sensitivity analysis (See Section~\ref{sec:tuningLambda} for details).
For a fair comparison, we selected the tuning parameter $s$ such that the numbers of genes selected using the GuidedSparseKmeans and the sparse K-means were close to the number of intrinsic genes in each simulation (i.e., 400). 

\begin{figure}[h!]
\centering
  \subfloat[Clustering results]{\label{fig:3(a)}
    \includegraphics[width=.45\textwidth]{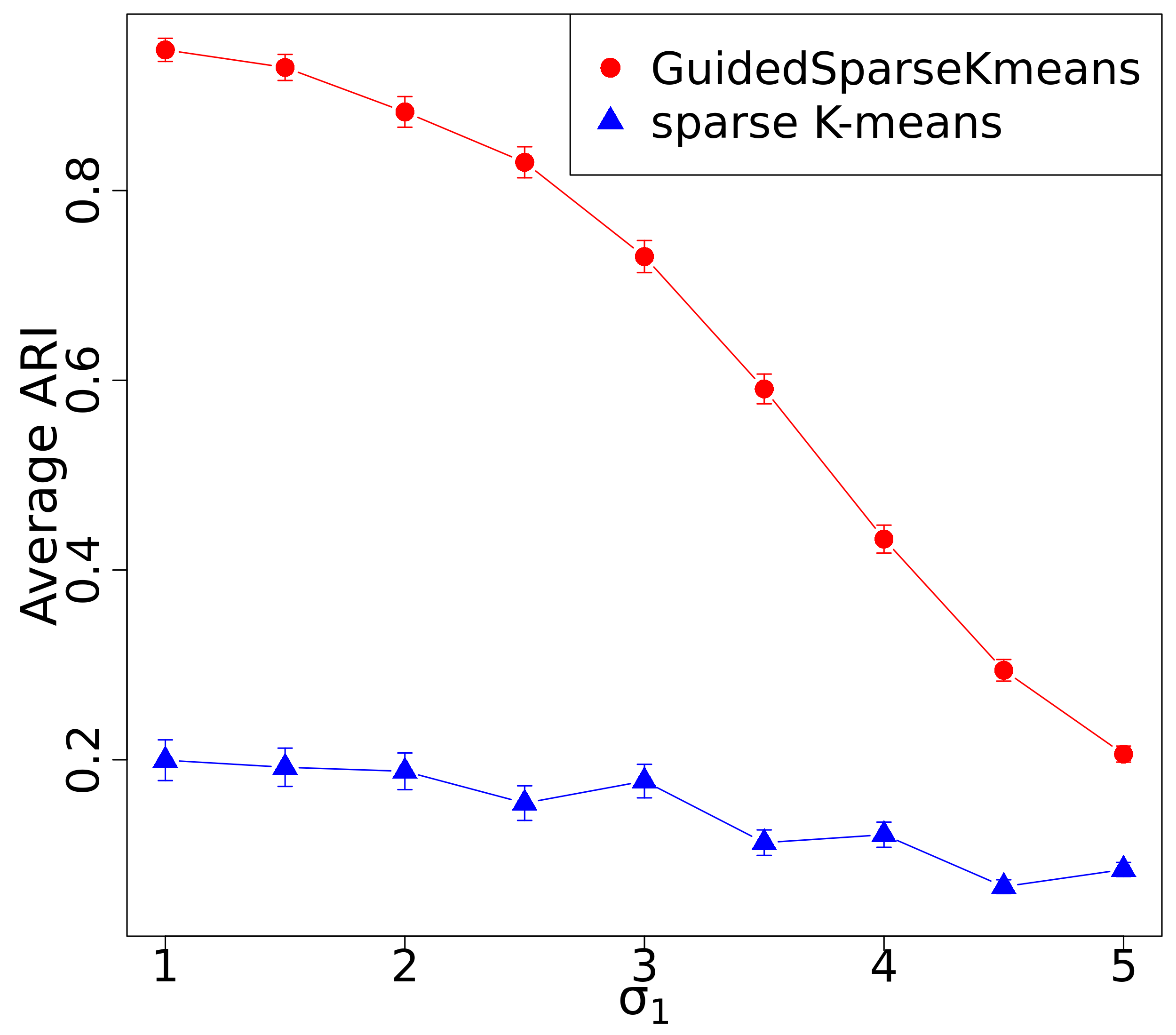}}
~
  \subfloat[Gene selection results]{\label{fig:3(b)}
    \includegraphics[width=.45\textwidth]{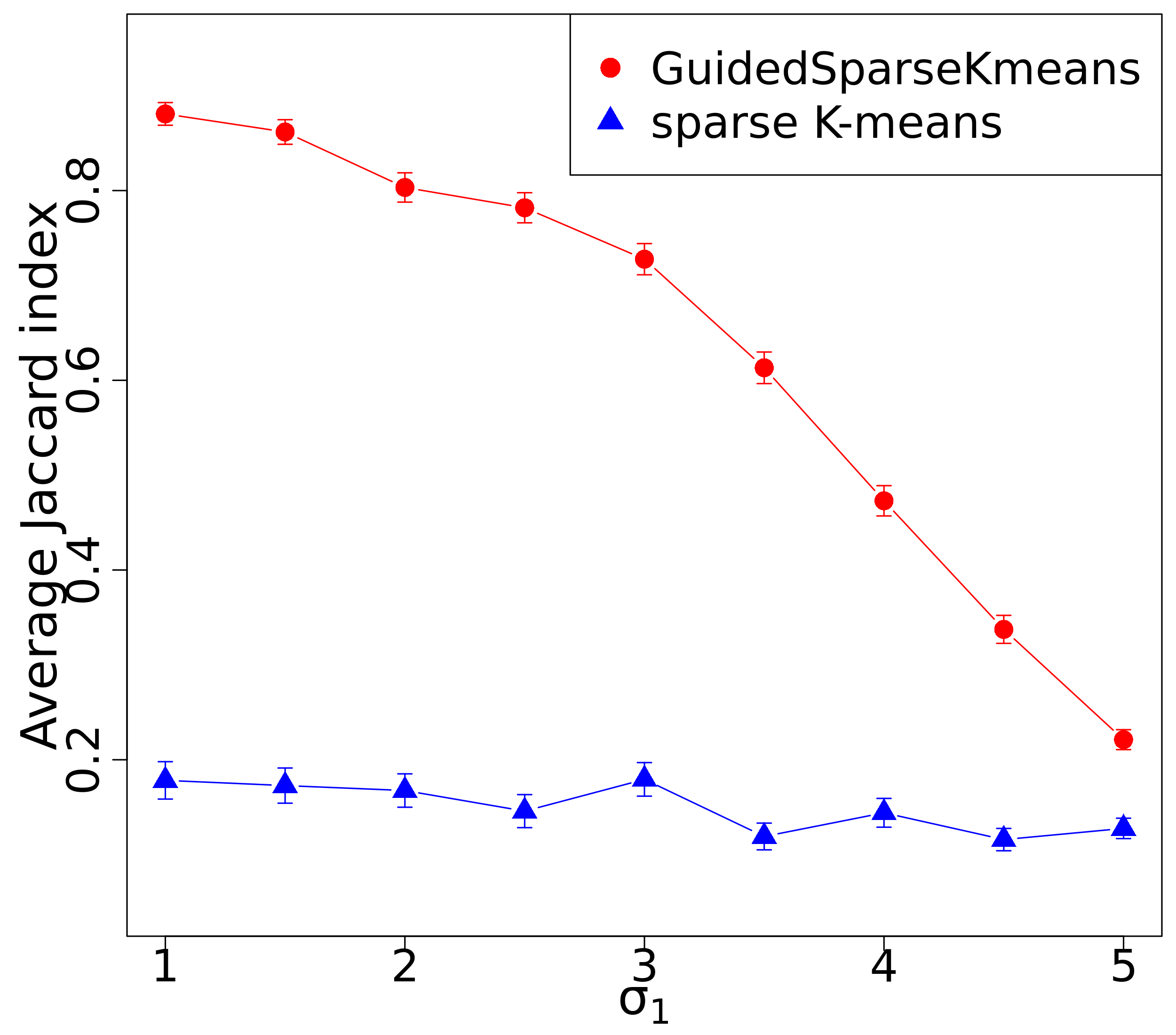}}
    
\caption{Simulation results for the comparison between the GuidedsparseKmeans and the sparse K-means with different values of biological variation $\sigma_1$. Mean estimates and standard errors were reported based on $B=100$ simulations. Figure 3(a): Clustering accuracy were evaluated using ARI. Figure 3(b): Gene selection accuracy were evaluated using Jaccard index.}
\label{fig:3}
\end{figure}

Table~\ref{tab:1} shows the comparison results of two methods with the biological variation $\sigma_1=3$. 
The GuidedSparseKmeans (mean ARI = 0.730) outperformed the sparse K-means (mean ARI = 0.178) in terms of clustering accuracy.
In addition, in terms of gene selection, 
the GuidedSparseKmeans (mean Jaccard index = 0.728) outperformed the sparse K-means (mean Jaccard index = 0.179).
To ensure the performance difference between the two methods was due to methodology itself instead of the specific tuning parameter selection, 
we also tried other number of selected genes (800 and 1,200) for all methods.
As shown in Table S1, the GuidedSparseKmeans still outperformed its competing method.
Further, we compared the gene selection performance in terms of the area under the curve (AUC) of a ROC curve, which aggregated the performance of all candidate tuning parameter $s$’s. 
And not surprisingly, the GuidedSparseKmeans (mean AUC = 0.881) achieved better performance than the sparse K-means (mean AUC = 0.590).
Further, under the settings with varying biological variation ($\sigma_1=1$ to 5 with interval 0.25),
Figure~\ref{fig:3(a)} and Figure~\ref{fig:3(b)} show that the GuidedSparseKmeans still outperformed the sparse K-means algorithm in terms of both clustering and gene selection accuracy,
though the performance of both methods decreased as the biological variation increased.

Additionally, we also compared the GuidedSparseKmeans with the sparse K-means under the settings with (i) varying clinical outcome variation, (ii) different number of true intrinsic genes, (iii) different number of confounding impacted genes, and (iv) different choice of number of clusters $K$. 
(i) By increasing the clinical outcome variation ($\sigma_2 = 6, 8,$ and $10$), we mimicked the decrease association of the disease subtypes and the clinical variable.
Table S2 shows that the performance of the GuidedSparseKmeans got worse when the clinical outcome variation increased.
This is not surprising because our algorithm relied on the clinical outcome variable to enhance feature selection. 
And the performance of the GuidedSparseKmeans still outperformed the performance of the  sparse K-means. 
(ii) By increasing number of true intrinsic genes (400, 800, and 1,200 genes), we mimicked larger signals in intrinsic genes. 
As shown in Table S3, the performance of both the GuidedSparseKmeans and the sparse K-means improved with increasing number of intrinsic genes.
And comparing these two methods, the GuidedSparseKmeans still outperformed the sparse K-means. 
(iii) By increasing number of confounding impacted genes (1,200, 1,600, and 2,000 genes), we mimicked more complicated confounding effect, and thus we expected it to be more difficult to identify the underlying disease subtypes.
As expected, the performance of both the GuidedSparseKmeans and the sparse K-means deteriorated with increasing number of confounding impacted genes (Table S4).
And comparing these two methods, the GuidedSparseKmeans still outperformed the sparse K-means. 
(iv) In simulation, the underlying number of clusters is 3. 
To examine the impact of misspecification of $K$, we varied the number of clusters $K = 2, 3, …, 6$. 
The result is shown in Table S5. 
$K = 3$ yielded the best performance in terms of clustering result and gene selection. When $K<3$ or $K>3$, the performance gradually decreased.
Thus, the misspecification of $K$ had great effects in the simulation since the 3 clusters in the simulations were clearly separated.

We further used cross-validation to examine the generalizability and replicability of the GuidedSparseKmeans in simulation.
We first simulated one dataset based on the basic simulation setting in Section~\ref{sec:simuSetting}.
Then we randomly divided the dataset into a training dataset (50\% of the total samples) and a testing dataset (50\% of the total samples). 
After applying the GuidedSparseKmeans on the training data to obtain selected genes, we used weighted K-means to get clustering results with the selected genes.
This procedure was repeated 50 times.
Table S6 shows the cross-validation evaluation results. The cross-validation ARI (0.712) was close to the mean ARI (0.730) in Table~\ref{tab:1}, which demonstrated that the generalizability and replicability of the GuidedSparseKmeans.

\begin{table}
    \caption{\label{tab:1}Comparison in the performance of clustering and gene selection between the GuidedSparseKmeans and the sparse K-means with biological variation $\sigma_1=3$. For a fair comparison, the number of selected genes were close to 400 for both methods. Mean estimates and standard errors were reported based on $B=100$ simulations. Clustering accuracy were evaluated by ARI. Gene selection accuracy were evaluated by Jaccard index and AUC.}
    \centering
    \fbox{%
    \begin{tabular}{c c c c c}
           & Clustering results & &\multicolumn{2}{c}{Gene selection results}\\
           & ARI & & Jaccard index & AUC \\
      \hline
      GuidedSparseKmeans  & 0.730 (0.017) & & 0.728 (0.016) & 0.881 (0.008)\\
      sparse K-means  & 0.178 (0.018) & & 0.179 (0.018) & 0.590 (0.010)\\
    \end{tabular}}
\end{table}

\subsection{METABRIC breast cancer example}
\label{sec:metabric}

In this section, we evaluated the performance of the GuidedSparseKmeans in the METABRIC breast cancer gene expression dataset \citep{curtis2012genomic}, which included samples from tumor banks in the UK and Canada.  
The METABRIC cohort contained gene expression profile of 24,368 genes and 1,981 subjects,
as well as various types of clinical outcome variables including 
the Nottingham prognostic index (NPI, continuous variable);
Estrogen receptor status (ER, binary variable);
HER2 receptor status (HER2, ordinal variable);
and overall survival.
We used each of these 4 phenotypic variables as the clinical guidance for the GuidedSparseKmeans, respectively. 
All datasets are publicly available at \url{https://www.cbioportal.org}. 
We matched the gene expression data to the clinical data by removing missing samples. 
After filtering out 50\% low expression genes based on the average expression level of each gene, 
we scaled the data such that each gene has mean expression value 0 and standard deviation 1. 
After these preprocessing procedures, we ended up with 12,180 gene features and 1,870 subjects. 

For each of these outcome variables, we calculated $U_g$ as the $R^2$ or pseudo-$R^2$ from the univariate regression between gene $g$ and an outcome variable,
where we used the linear model for continuous outcomes; generalized linear models for binary and ordinal outcomes; and the Cox regression model for survival outcomes. 
We set the tuning parameter $K=5$ based on the gap statistics analysis, which is consistent with acknowledgement that there are 5 types of breast cancer by the PAM50 definition \citep{parker2009supervised}.
The tuning parameter $\lambda$ of GuidedSparseKmeans was selected automatically (See methods in Section~\ref{sec:tuningLambda}).
We selected the tuning parameter $s$ such that the number of selected genes was closest to 400.     
To benchmark the performance of the GuidedSparseKmeans,
we further compared with the non-outcome-guided clustering algorithm (the sparse K-means) and the two-step clustering method,
with  
number of selected genes around 400 (See Table~\ref{tab:2}). 
Fixing similar number of selected genes for different methods will minimize the influence resulted from unequal number of selected genes. and emphasize the comparisons of methodologies themselves.

\subsubsection{Evaluating the clustering performance }

\begin{figure}[h!]
\centering
  \subfloat[Heatmap from the NPI-GuidedSparseKmeans]{\label{fig:4(a)}
    \includegraphics[width=.32\textwidth]{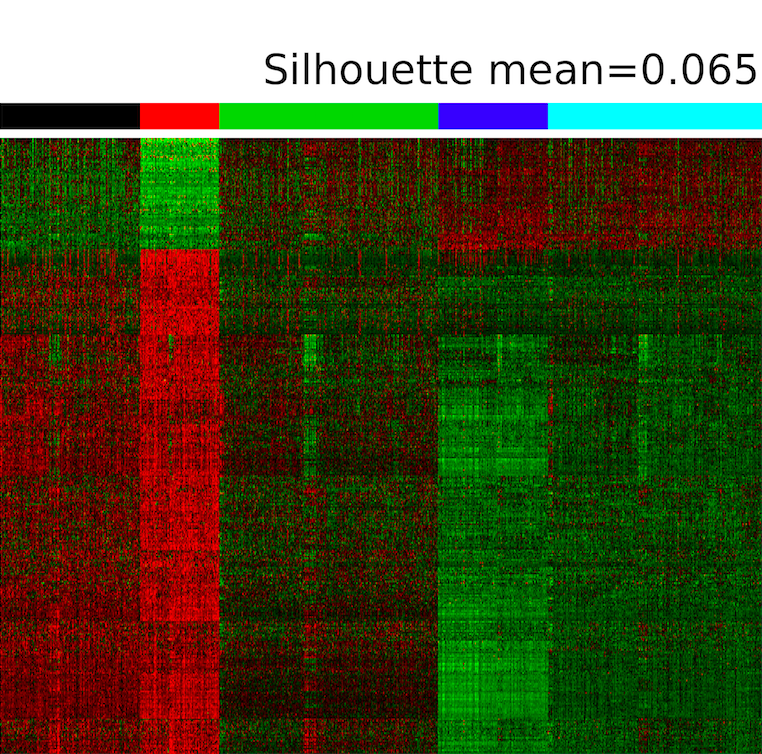}}
~
  \subfloat[Heatmap from the ER-GuidedSparseKmeans]{\label{fig:4(b)}
    \includegraphics[width=.32\textwidth]{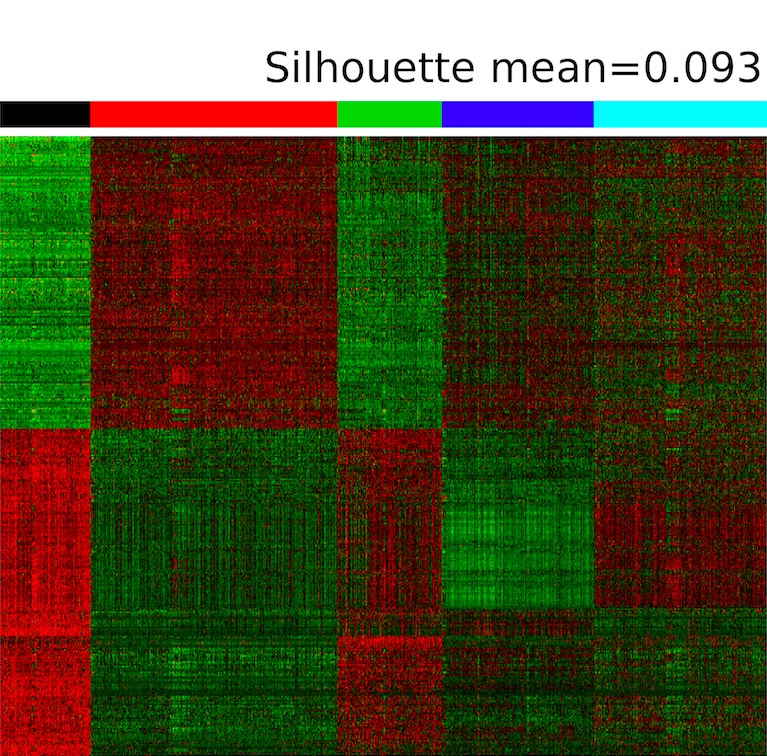}}
~
  \subfloat[Heatmap from the HER2-GuidedSparseKmeans]{\label{fig:4(c)}
    \includegraphics[width=.32\textwidth]{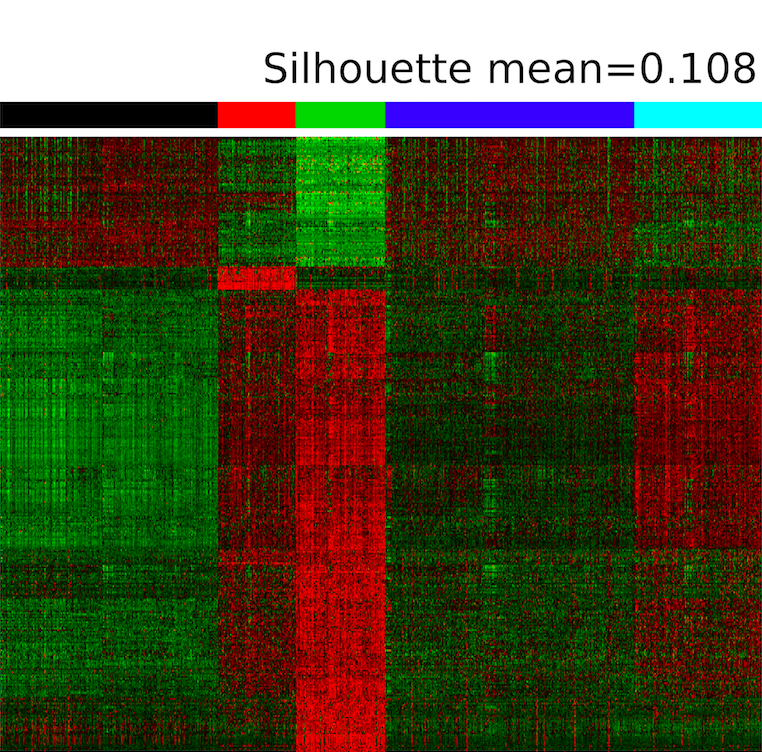}}
    
  \subfloat[Survival curves from the NPI-GuidedSparseKmeans]{\label{fig:4(d)}
    \includegraphics[width=.32\textwidth]{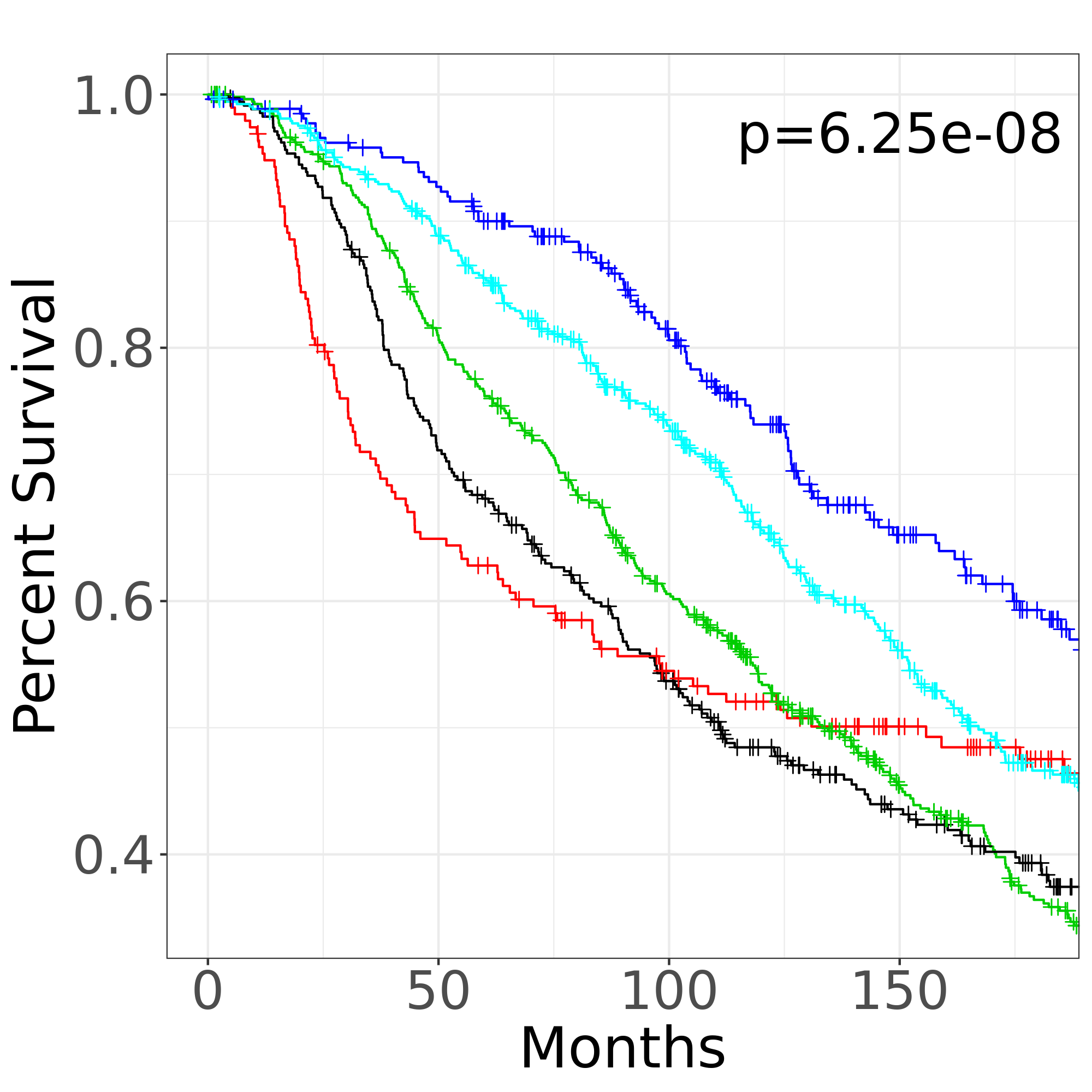}}
~
  \subfloat[Survival curves from the ER-GuidedSparseKmeans]{\label{fig:4(e)}
    \includegraphics[width=.32\textwidth]{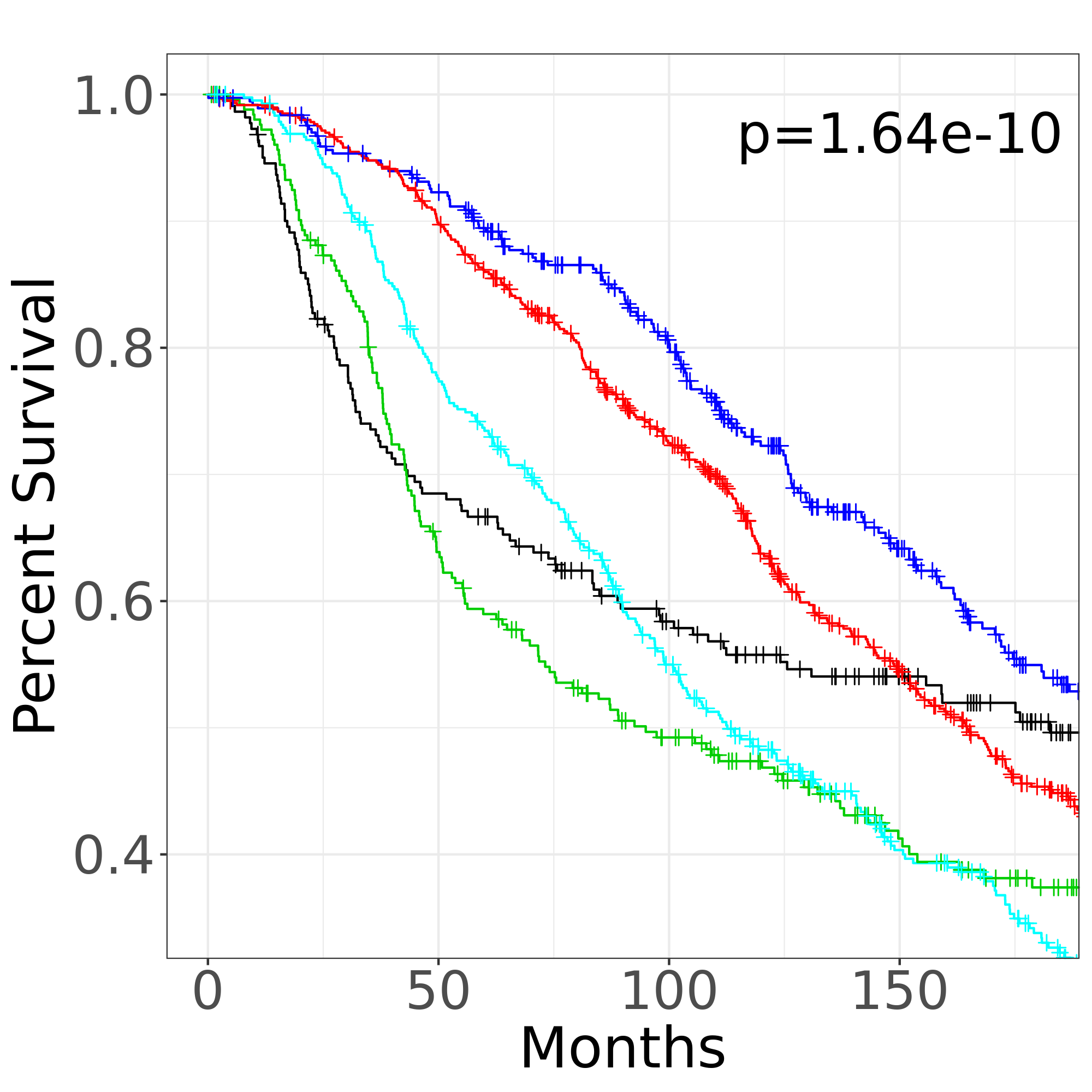}}
~
  \subfloat[Survival curves from the HER2-GuidedSparseKmeans]{\label{fig:4(f)}
    \includegraphics[width=.32\textwidth]{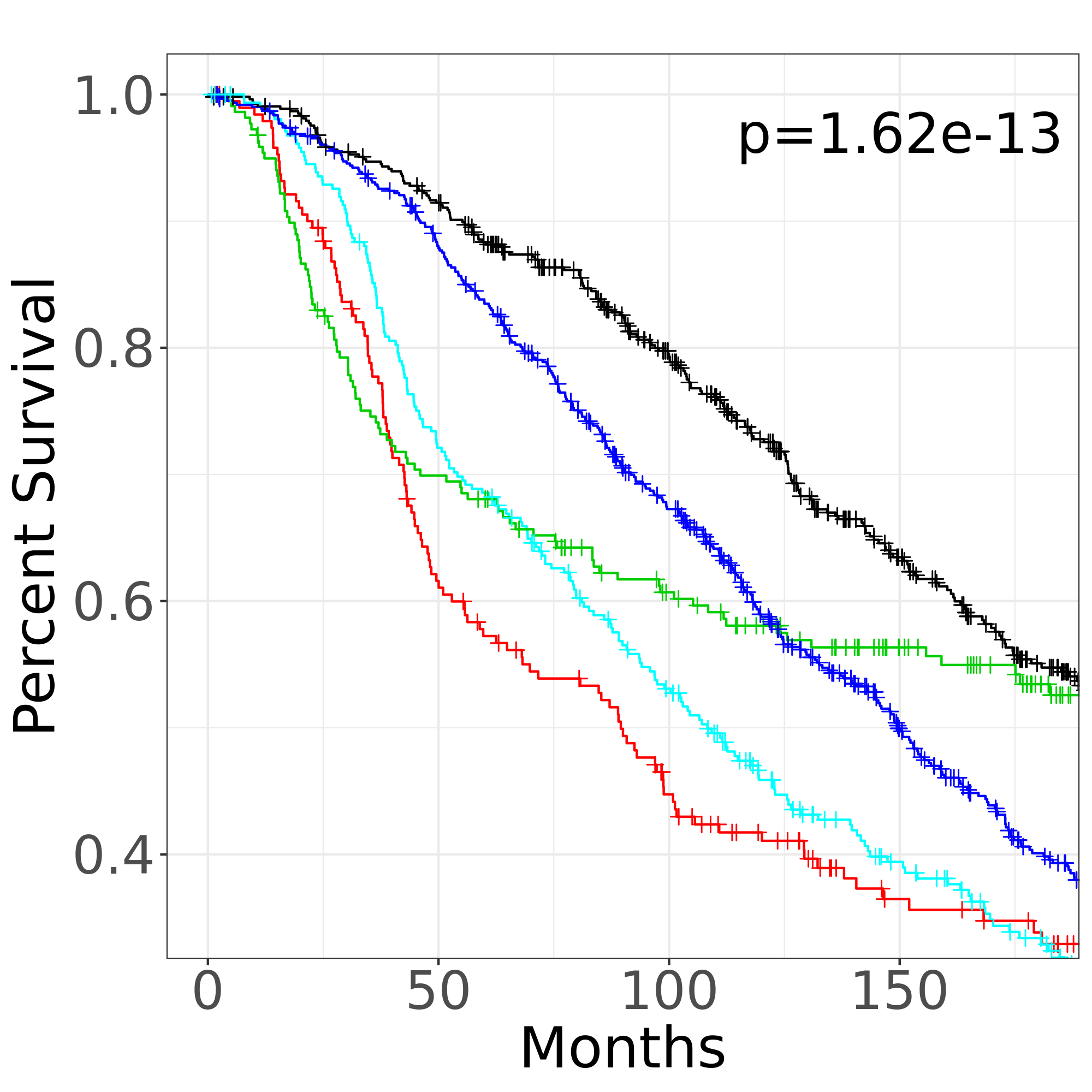}}

\caption{Gene expression heatmap and Kaplan–Meier survival curve of the METABRIC dataset using the NPI-GuidedSparseKmeans, the ER-GuidedSparseKmeans and the HER2-GuidedSparseKmeans. In heatmap (a)(b)(c), rows represent genes and columns represent samples. Red color represents higher expression and green color represents lower expression. The samples are divided into 5 clusters with 5 colors in the bar above the heatmaps. ``Silhouette mean'' is the mean Silhouette score of the selected genes, with larger value indicating better separation between clusters. In survival curves (d)(e)(f), the color of the survival curve for each subtype is corresponding to the subtype color in the heatmap. The p-values for the survival difference is marked on the top-right corner. Note that these significant p-values might be biased as survival is partly correlated with the clinical variable used.}
\label{fig:4}
\end{figure}

Since we did not have the underlying true clustering result, we used the survival difference between subtypes to examine whether the resulting subtypes were clinically meaningful. As shown in Figure~\ref{fig:1}, Figure~\ref{fig:4}, and Table~\ref{tab:2}, 
the Kaplan-Meier survival curves of the five disease subtypes obtained from each GuidedSparseKmeans method or the two-step method were well separated with significant p-values,
but not for the non-outcome-guided method (the sparse K-means). 
This is not unexpected because except for the sparse K-means method, 
all the other methods have guidance from clinical outcomes.
Remarkable, even though the NPI-GuidedSparseKmeans, the ER-GuidedSparseKmeans, and the HER2-GuidedSparseKmeans did not utilize any survival information, 
they still achieved good survival separation.
Probable reason could be that these clinical outcome variables were all related to the disease, and thus could impact the overall survival. 

We further compared the clustering results obtained from each method with the PAM50 subtypes, which were considered as the gold standard for breast cancer subtypes. 
Table~\ref{tab:2} shows that the ARI values from four types of GuidedSparseKmeans (0.237 $\sim$ 0.292) were greater than the ARI values from the sparse K-means (0.017) or the two-step method (0.177), indicating that the subtype results from the GuidedSparseKmeans were closer to the gold standard than those from the sparse K-means or the two-step method.
These results are expected because the GuidedSparseKmeans could utilize the clinical outcome variables to obtain biologically interpretable clustering results.

For the heatmap patterns, in general the sparse K-means achieved the most homogenous clustering results (mean Silhouette = 0.099), followed by the GuidedSparseKmeans (mean Silhouette = 0.065 $\sim$ 0.108), and the two-step method achieved the most heterogenous clustering results (mean Silhouette = 0.058).  
Notably, the HER2-GuidedSparseKmeans achieved even better mean Sillhouette score than the spares K-means (mean Silhouette = 0.108).  

To examine whether the subtype clusters from the GuidedSparseKmeans were relevant to the guidance of the clinical outcome variables, 
we calculated the relevancy score by Equation~\ref{eq:relevancy}.
Not surprisingly, the ER-GuidedSparseKmeans achieved the highest relevancy score (0.764), followed by NPI-GuidedSparseKmeans (0.690), Survival-GuidedSparseKmeans (0.559), and HER2-GuidedSparseKmeans (0.305) indicating the clustering results from the GuidedSparseKmeans with each of four clinical outcomes reasonably reflected the information of their clinical outcomes.

To ensure the performance difference between the two methods was due to methodology itself instead of the specific tuning parameter selection, 
we also tried other targeted number of selected genes (800, 1,200 and 1,600) for all methods.
As shown in Table S7 the GuidedSparseKmeans still outperformed its competing methods.

Collectively, as discussed in the motivating example, 
the sparse K-means algorithm only sought for homogeneous subtype clustering patterns, regardless of whether the resulting subtypes were clinically meaningful.
The two-step method only sought for subtypes that were clinically meaningful, which could result in high inter-individual variabilities within each subtype cluster.
And the GuidedSparseKmeans sought for both homogeneous subtype clustering patterns and relevant clinical interpretations.

\begin{table}
    \caption{\label{tab:2}Comparison in clustering and gene selection of the four GuidedSparseKmeans methods, the sparse K-means method, and the two-step method. The number of genes selected by each method was close to 400. The relevancy of the subtype clustering result and the clinical outcome variable was evaluated by the relevancy score. The clustering results were compared with PAM50 subtype in terms of ARI. The separation between resulting clusters and cohesion in respective clusters were evaluated by the Silhouette score. P-values of survival differences of identified subgroups were calculated based on the log-rank test.}
    \centering
    \fbox{%
    \begin{tabular}{c| c c c c c c c}
      Method &Guidance & Genes &Relevancy &ARI &Silhouette &p-value &Time\\
      \hline
              &NPI(continuous) &399 &0.690 &0.237 &0.065 &$6.25\times10^{-8}$ &31 secs\\
      Guided  &ER(binary)      &411 &0.764 &0.292 &0.093 &$1.64\times10^{-10}$ &3.3 mins\\
      SparseKmeans &HER2(ordinal) &394 &0.305 &0.292 &0.108 &$1.62\times10^{-13}$ &19.2 mins\\
              &Survival        &384 &0.559 &0.244 &0.073 &$5.33\times10^{-10}$ &2.9 mins\\
      \hline
      sparse K-means &         &408 &      &0.017 &0.099 &0.072 &9.8 mins\\
      \hline
      two-step method &Survival &400 &     &0.177 &0.058 &$2.92\times10^{-19}$ &2.5 mins\\
    \end{tabular}}
\end{table}

\subsubsection{Evaluating the gene selection performance }

\begin{figure}[h!]
\centering
\includegraphics[scale=0.35]{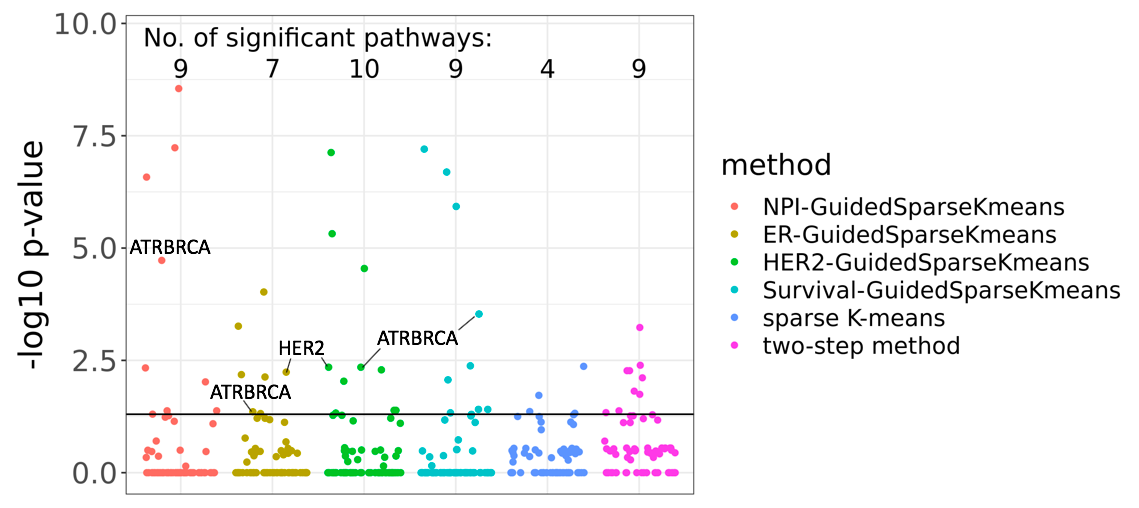}
\caption{Pathway enrichment analysis results for the GuidedSparseKmeans with four different types of guidance, the sparse K-means, and the two-step method. The number of significant pathways ($p < 0.05$) for each method is on the top.}
\label{fig:5}
\end{figure}

To evaluate whether the selected genes were biologically meaningful, we performed pathway enrichment analysis via the BioCarta pathway database using the Fisher's exact test. 
Figure~\ref{fig:5} shows the jitter plot of p-values for all pathways obtained from each method.
Using $p = 0.05$ as significance cutoff, 
the points above the black horizon line are the significantly enriched pathways without correcting for multiple comparisons.
We observed that the genes selected from the GuidedSparseKmeans (number of significant pathways = 7 $\sim$ 10) and the two-step method (number of significant pathways = 9) had better biological interpretation than the sparse K-means method (number of significant pathways = 4). 
Also, the top p-values from the GuidedSparseKmeans were lower than those from the two-step method.
Notably, the genes selected by the ER-GuidedSparseKmeans and the HER2-GuidedSparseKmeans were enriched in HER2 pathway, 
with $p = 5.78\times10^{-3}$ and $5.14\times10^{-3}$, respectively. 
This is in line with previous research that ER signaling pathway and HER2 signaling pathway are closely related to breast cancer \citep{giuliano2013bidirectional}.
In addition, the genes selected by the GuidedSparseKmeans with each of four clinical outcomes were enriched in ATRBRCA pathway, 
which has been found closely related to breast cancer susceptibility \citep{paul2014breast}.
However, these hallmark pathways of breast cancer were not picked up by the sparse K-means algorithm or the two-step method, and only the GuidedSparseKmeans selected the most biologically interpretable genes.

\subsubsection{Evaluating the computing time}

The computing time for the GuidedSparseKmeans in this real application was 31 seconds for continuous outcome guidance, 
2.9 minutes for survival outcome guidance, 
3.3 minutes for binary outcome guidance, 
and 19.2 minutes for ordinal outcome guidance.
Since our high-dimensional expression data contains 12,180 genes and 1,870 samples, 
such computing time is reasonable and applicable in large transcriptomic applications.

In general, the computing time of our method was faster than the sparse K-means method (9.8 minutes), except for the ordinal outcome guidance. 
This was because we used $U_g$ to guide the initialization of the weights instead of equal weight initialization in the sparse K-means,
which saved lots of time in the first step weighted K-means iteration.  
The ordinal-outcome-GuidedSparseKmeans was slow, because of the slow fitting of ordinal logistic regression models. 
Our methods were generally slower than the two-step method, except for the continuous outcome guidance.
This was because the two-step method only needed to update the clustering assignment once, 
while ours needed to iteratively update the clustering assignment and gene weight. 
The continuous-outcome-GuidedSparseKmeans was super-fast (31 seconds), because the $R^2$ of univariate linear models can be efficiently calculated as the square of the Pearson correlation between a gene and the outcome variable.

\subsubsection{Model sensitivity and generalizability}

In the optimization procedure (Section~\ref{sec:opt}), we chose $r=400$ as the initial number of selected genes. 
To examine the impact this initial number $r$ on the clustering results, we performed sensitivity analysis by varying $r = 400, 600, 800$.
As shown in Table S8, the varying choice of $r$ did not seem to have large impact the clustering results and the feature selection result.

To examine the impact of choices of number of clusters, we varied the number of clusters $K = 3, 4, …, 7$. 
The result is shown in Table S9. 
The performance of different $K$’s were similar, and there was not a $K$ that could simultaneously achieve the best relevancy score, ARI with PAM 50, Silhouette score, and survival p-value.
The choices of $K$ had greater effects on the simulation than the real data application since the clusters in the simulations were clearly separated, while in the real data and the difference between clusters ere gradual instead of steep.

Further, we used cross-validation to evaluate the generalizability and the replicability of the results from the GuidedSparseKmeans.
To be specific, we randomly split the data (n = 1,870) into a training dataset (n = 935) and testing dataset (n = 935).
We applied the GuidedSparseKmeans algorithm on the training dataset and obtain the selected genes. 
Then we used the selected genes to predict the clustering result in the testing dataset.
This procedure was repeated 50 times, and the mean ARI (compared to PAM) or median survival p-values were reported.
As shown in Table S10, the cross-validation clustering accuracy (compared to PAM) for the NPI-GuidedSparseKmeans, the ER-GuidedSparseKmeans, the HER2-GuidedSparseKmeans, and the Survival-GuidedSparseKmeans were 0.265, 0.296, 0.296, and 0.212, respectively, which were close to the mean ARI 0.237, 0.292, 0.292, and 0.244 in Table~\ref{tab:2}. 
The survival p-values from the cross-validation were still very significant, though they were a little bit larger than the whole dataset, which was because we only used half the sample size. 
These results reveal that the result from the GuidedSparseKmeans are generalizable and replicable.

\subsection{Alzheimer disease example}
\label{sec:Alzheimer}

In this section, we applied the sparse K-means and the GuidedSparseKmeans to an RNA-seq dataset of Alzheimer disease (AD). 
AD is a devastating neurodegenerative disorder affecting elder adults, with neurofibrillary tangles and neuritic plaques as disease hallmarks. 
The dataset we used included 30,727 transcripts from post mortem fusiform gyrus tissues of 217 AD subjects, which is available at \url{https://www.ncbi.nlm.nih.gov} under GEO ID GSE125583 \citep{srinivasan2020alzheimer}.
After normalizing the gene expression data, filtering out 50\% low expression genes based on the average expression level of each gene, and scaling each gene to mean 0 and standardization 1, we ended up with 15,363 gene features. 
We utilized the Braak stage as the clinical outcome variable, which is a semiquantitative measure of severity of neurofibrillary tangles \citep{braak1991neuropathological}. Since the Braak system has 6 stages (1 $\sim$ 6), we treated it as a continuous variable, 
and calculated $U_g$ as the $R^2$ from the univariate linear regression between gene $g$ and the Braak stage. We set the tuning parameter $K=6$ to be consistent with the number of Braak stages. The tuning parameter $\lambda$ of GuidedSparseKmeans was selected automatically using the method in Section~\ref{sec:tuningLambda}. The tuning parameter $s$ was selected such that the number of selected genes was 396, which was closest to 400.  
The computing time for this example was only 7 seconds. 
Also, the sparse K-means was applied to obtain about 400 genes and the computing time was 22 seconds.

\begin{figure}[h!]
\centering
  \subfloat[Heatmap from the Sparse K-means]{\label{fig:6(a)}
    \includegraphics[width=.45\textwidth]{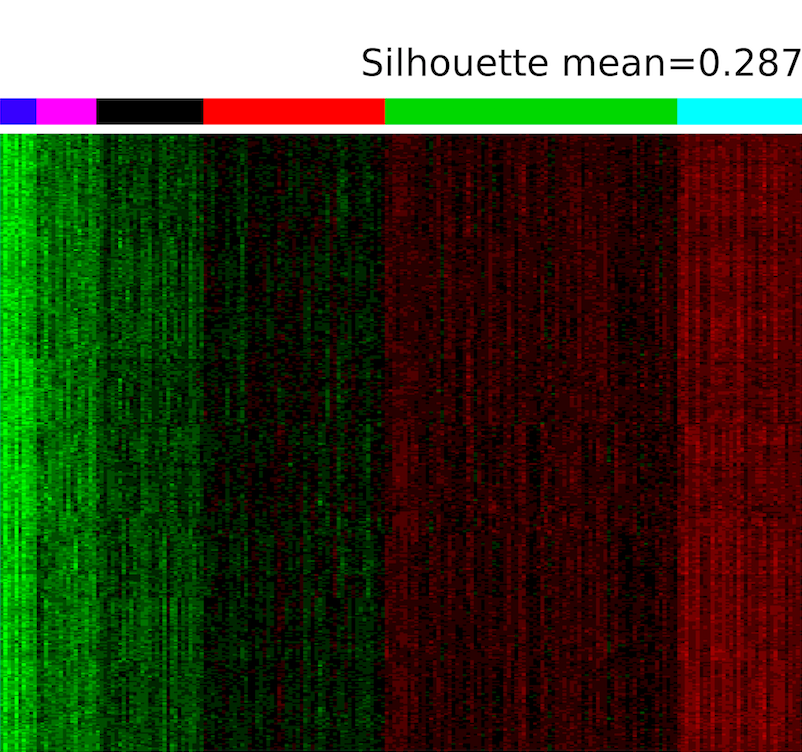}}
~
  \subfloat[Heatmap from the Braak-GuidedSparseKmeans]{\label{fig:6(b)}
    \includegraphics[width=.45\textwidth]{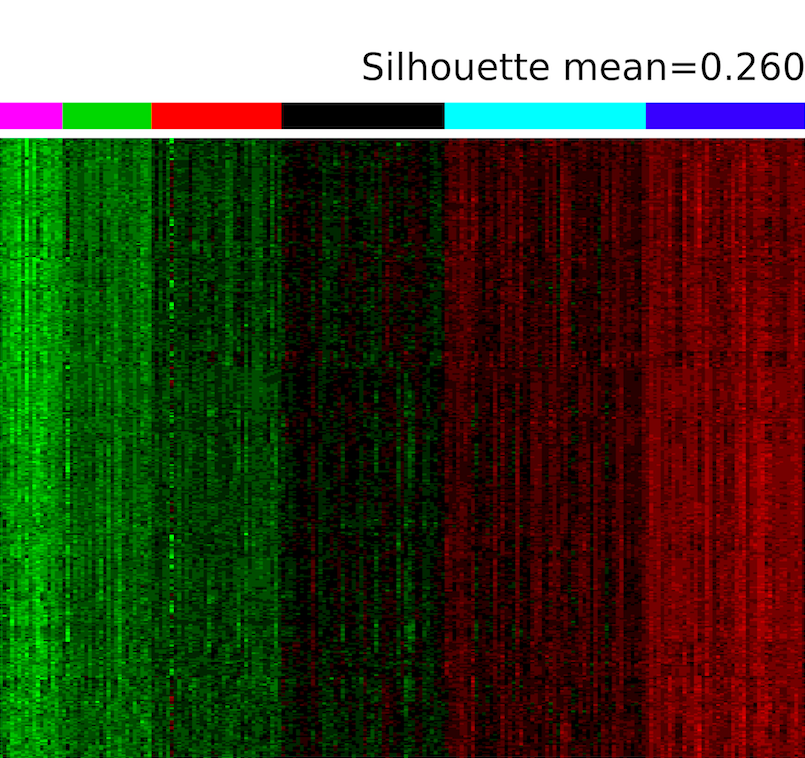}}

\caption{Gene expression heatmap of the Alzheimer's disease dataset using the sparse K-means and the Braak-GuidedSparseKmeans. In the heatmap, rows represent genes and columns represent samples. Red color represents higher expression and green color represents lower expression. The samples are divided into 6 clusters with 6 colors in the bar above the heatmap.}
\label{fig:6}
\end{figure}

The resulting heatmap of the selected genes from the sparse K-means and the GuidedSparseKmeans in Figure~\ref{fig:6} showed a gradient effect from left (green color, lower level of expression) to right (red color, higher level of expression).
We labeled the clusters using the integers 1 to 6 which corresponds to the gene expression level from low to high. 
To evaluate whether the resulting AD subtypes were biologically meaningful, we calculated the Pearson correlation between the subtype labels and the Braak stages.
The correlation coefficients were 0.056 and 0.282 for the sparse K-means and the GuidedSparseKmeans respectively, indicating that the AD subtypes obtained by the GuidedSparseKmeans were more related to neurofibrillary tangles (a key hallmark of AD) than those obtained by the sparse K-means.
These results are expected since the GuidedSparseKmeans could take advantage of the clinical responses.

To examine whether clusters from the GuidedSparseKmeans were relevant to the guidance of the clinical outcome variable Braak staging, 
we calculated the relevancy score by Equation~\ref{eq:relevancy}. 
The relevancy score was high (Rel = 0.309), which is not unexpected because the GuidedSparseKmeans reflected the information of the clinical outcome.  
To further examine whether the genes selected from the GuidedSparseKmeans were biologically meaningful, we performed pathway enrichment analysis via the BioCarta pathway database using the Fisher's exact test. Notably, the genes selected by the Braak-GuidedSparseKmeans were enriched in GPCR signaling pathway ($p = 6.59\times10^{-6}$) and NOS1 signaling pathway ($p = 1.61\times10^{-5}$), both of which were involved in the neuropathological processes of AD \citep{zhao2016g, reif2011association}. 
However, these hallmark pathways were not picked up by the sparse K-means method.
Therefore, compared with the sparse K-means, the GuidedSparseKmeans successfully identified AD subtypes and the intrinsic genes that were both closely related to AD.

\section{Discussion}

Since our method can only accommodate one clinical outcome variable,
one issue is how to properly select the clinical guidance given multiple available clinical outcomes. 
We suggest to select the outcome variable of the most biological interest as clinical guidance, based on domain knowledges. Taking the Alzheimer's disease as an example, the variable Braak stage was chosen due to its important role of measuring the severity of neurofibrillary tangles in AD.    
If no domain knowledge can be used or one prefers a data driven approach, we suggest to try several outcome variables as guidance,
and decide the best outcome guidance based on the biological interpretability of the clustering results (i.e., survival difference, pathway analysis). 
For example, in the breast cancer application, we would recommend the clustering result obtained by the HER2-GuidedSparseKmeans since it led to both the most significant survival difference and meaningful pathway analysis result.   
To further address this issue,
we will develop a multivariate outcome-guided sparse K-means in the future, which will incorporate multiple outcomes variables as the clinical guidance.

Properly specifying the disease-related clinical outcomes is essential for the disease subtyping.
If the guidance term (clinical outcome) has no association with the underlying disease subtype, the result from the GuidedSparseKmeans may further deviate from the true subtypes.
However, we still think our algorithm has its unique merit because many hallmark clinical variables have been established for many diseases (e.g., ER for breast cancer, braak stage for Alzheimer’s disease).
With the advancement of biomedical research, more disease-related clinical outcomes will be discovered, and our algorithm will become more applicable. 

In this manuscript, we estimated the tuning parameters of the GuidedSparseKmeans $K$, $\lambda$, and $s$ by gap statistics, sensitivity analysis, and extended gap statistics, respectively. 
Other approaches that can simultaneously estimate tuning parameters were recently proposed \citep{li2019simultaneous}.
One future direction is to develop methods to jointly estimate these tuning parameters.
We also considered the effects of the number of selected genes and mis-specification of the number of clusters $K$ on clustering and gene selection performance of the GuidedSparseKmeans. 
The GuidedSparseKmeans still outperformed the sparse K-means regardless of the number of selected genes in simulations and breast cancer application, respectively. 
In simulation, the mis-specification of $K$ had large impacts on the clustering performance of the GuidedSparseKmeans, while in the breast cancer application, the performance of the GuidedSparseKmeans was slightly affected by different $K$. This is reasonable since the simulation datasets have clear and direct clusters but in the real data and the difference between clusters are gradual instead of steep.

In summary, 
we proposed an outcome-guided sparse K-means algorithm, 
which is capable of identifying subtype clusters via integrating clinical data with high-dimensional omics data. 
Our innovative methodology simultaneously solves many statistical challenges, 
including gene selections from high-dimensional omics data; 
incorporations of various types of clinical outcome variable (e.g., continuous, binary, ordinal, count or survival data) as guidance;
automatic selections of tuning parameters to balance the contribution between the intrinsic genes and the clinical outcome variable; 
and evaluations of the relevancy of the resulting subtype clusters and the outcome variable. 
The superior performance of our method was demonstrated in simulations,
cancer application (breast cancer gene expression data),
and complicated neurodegenerative disease (Alzheimer’s disease). 
The computing time for the GuidedSparseKmeans was fast, especially for the continuous outcome guidance, with 31 seconds for the breast cancer application with 12,180 genes and 1,870 samples, 
and 7 seconds for the Alzheimer’s disease application with 15,363 genes and 217 samples.
Our method has been implemented in R package ``GuidedSparseKmeans'', which is available at GitHub (\url{https://github.com/LingsongMeng/GuidedSparseKmeans}).
With the accumulation of rich clinical data and omics data in public data repositories, 
we expect our innovative and fast method can be very applicable in identifying disease phenotype related subtypes.

\section{Acknowledgement}

L.M., D.A., and Z.H. are partially supported by NIH 2R01AI067846, G.T. is partially supported by NIH R01CA190766, R01MH111601, and R21LM012752.

\section{Data Availability Statement}

The METABRIC breast cancer gene expression dataset and the Alzheimer disease RNA-seq dataset that support the findings of this study are openly available at \url{https://www.cbioportal.org}, and \url{https://www.ncbi.nlm.nih.gov} under GEO ID GSE125583, respectively.

\bibliographystyle{rss.bst}
\bibliography{references.bib}
\end{document}